\documentclass[letterpaper]{JHEP3}
\usepackage{graphicx}
\usepackage{psfrag}
\usepackage{url}
\usepackage{units}
\usepackage{amsmath}

\providecommand{\MSbar }{\ensuremath{ \overline{\rm MS} }}
\newcommand \T[1]{\vec{#1}_T}

\newcommand\VCbox[1]{%
   \raisebox{-0.5\height}{#1}%
}

\makeatletter 
\preprint{
\@date
}
\makeatother

\title{Initial State Parton Showers Beyond Leading Order}

\author{John C. Collins\\
        Physics Department,
        Penn State University,\\
        104 Davey Laboratory,
        University Park PA 16802-6300, U.S.A. \\
        E-mail: \email{collins@phys.psu.edu}}
\author{Xiaomin Zu \\
        High Energy Physics\\
        505 Keen Building
        Tallahassee, FL32306-4350, U.S.A. \\
        E-mail: \email{xiaomin@hep.fsu.edu}
}
\date{11 April 2005}

\abstract{%
  We derive a new method for initial-state collinear showering in
  Monte-Carlo event generators which is based on the use of
  unintegrated parton correlation functions.  Combined with a
  previously derived method for final-state showering, the method
  solves the problem of treating both the hard scattering and the
  evolution kernels to be used in arbitrarily non-leading order.
  Although we only treat \emph{collinear} showering, so that further
  extensions are needed for QCD, we have discovered several new
  results: (1) It is better to generate exact parton kinematics in the
  hard scattering rather than with the subsequent parton showering,
  and similarly at each step of the showering.  (2) Parton showering
  is then done 
  conditionally on the exact energy-momentum of the initiating parton.  (3)
  We obtain a factorization for structure functions in terms of parton
  correlation functions so that parton kinematics can be treated
  exactly from the beginning.  (4) We obtain two factorization
  properties for parton correlation functions, one in terms of
  ordinary parton densities and one, suitable for event generation, in
  terms of parton correlation functions themselves.
}


\begin{document}

%==========================================================
\section{Introduction}
\label{sec:intro}

Monte-Carlo event generators (MCEGs) provide a very powerful approach
to implementing perturbative QCD to predict many observables in hard
processes.
However, despite a considerable amount of work --- see the paper by
Frixione and Webber \cite{NLO.MC} and the references reviewed there
--- current algorithms do not systematically extend beyond an improved
leading-logarithm approximation.  This is in contrast to factorization
theorems for hard inclusive processes, where it is known how both
hard-scattering coefficients and evolution kernels may be computed, in
principle, to arbitrarily high order in perturbation theory.  As a
computational tool, event generators not only provide estimates of the
exclusive components of cross sections, but they have the important
property that event generation needs a time roughly linear
in the number of particles.  In contrast, the computation of Feynman graphs
for cross sections involving $N$ particles needs resources
proportional to roughly $N!$.

In this paper, we provide a solution to the problem of non-leading
corrections in MCEGs when initial-state collinear showering is
involved.  Combined with a method one of us developed \cite{phi3} for
final-state collinear showering, this allows a complete treatment for
deep-inelastic scattering, for example.  Our treatment in this paper
is restricted to collinear showering, so it applies as it stands only
to non-gauge theories.  However, even though substantial extension of
the method will be needed to handle QCD, with its soft gluon emission,
our work solves a number of important conceptual problems whose
solution is also needed in QCD.

We will give a more detailed account of the problems we are concerned
about in Sec.\ \ref{sec:rationale}.  Here we will simply remark that
the algorithms in standard MCEGs are based on the factorization
theorem \cite{factn} for inclusive hard processes, whereas MCEGs
concern themselves with the exclusive components of cross sections.
Thus the standard factorization theorem does not really provide a
fully satisfactory foundation for MCEGs; a more general and powerful
theorem is needed.  This can be seen in several ways:
\begin{itemize}
\item The proofs \cite{factn} of standard factorization explicitly
  involve sums over unobserved final states, but MCEGs resolve the full
  final state.
\item Approximations on the kinematics of internal partons are made
  that become invalid if the exact final state is observed: the
  approximations violate exact momentum conservation.  Moreover, as we
  will show in Sec.\ \ref{sec:rationale}, the approximations interfere
  with the factorized structure of probabilities (or weights) that
  enables the stochastic generation of events to take a time linear in
  event size.
\item In standard factorization, cross sections that are infra-red and
  collinear safe are approximated by singular distributions like
  \begin{equation}
  \label{eq:singular}
    (1+\alpha_sA)\, \delta(x) + \alpha_sB \left( \frac{1}{x} \right)_+ + \ldots .
  \end{equation}
  While such approximations can be valid \emph{when integrated with a
    smooth function}, they are not valid when a fully differential
  cross section is under consideration.
\item Although the previous problem can be alleviated by some kind of
  resummation, it is preferable to have an improved factorization
  theorem from the beginning.
\end{itemize}
The most fundamental part of the proof of factorization, the analysis
of the momentum-space regions that give leading contributions to the
cross section, is indeed generally applicable, and so it does
give a correct starting point for deriving algorithms for MCEGs.
What is at issue are the further steps, and in particular the
approximations, that are used to obtain the standard form of
factorization; these approximations are only appropriate when there is an
inclusive sum over the hadronic final state.

Our approach solves the problems by working with parton correlation
functions that are defined without an integral over parton kinematics,
unlike conventional parton densities.  This avoids the need to use
approximated parton kinematics in places where the approximation is
invalid.  Moreover, we will also work with parton correlation
functions and cross sections that are defined without an integral over
hadronic final states (or, more conveniently, with an arbitrary weight
function of the hadronic final state).  This gives us the
appropriate tools for systematically and precisely treating the actual
physical situations which MCEGs aim to approximate.

Of course, parton correlation functions are more complicated objects
than parton densities, so without further information a lot of
predictive power is lost.  We solve this issue with extra
factorization theorems for the parton correlation functions at large
parton virtuality; the showering algorithm is what implements these.

At low parton virtuality, the information in parton correlation functions
corresponds to the information in non-perturbative hadronization
models that are present in conventional MCEGs.  There is no gain or
loss of predictive power here.

We will find that we can explicitly represent all probabilities and
cross sections for a MCEG in terms of integrals over parton
correlation functions, and hence in terms of explicit operator matrix
elements.  The systematic extension to non-leading order and
non-leading logarithms then becomes straightforward, with the use of
the subtractive methods advocated by us in \cite{phi3,jcc1}.

Related work is in papers by Watt, Martin and Ryskin \cite{MRW1,MRW2}
who have proposed the use of what they call ``doubly unintegrated
parton distributions'' (DUPDFs).  This concept  coincides with our
``parton correlation functions''.  Their primary motivation is the
same as ours: the need to treat parton kinematics exactly in certain
observables in a way that is fully compatible with ordinary parton
densities.  Their work treats QCD, so it is of direct phenomenological
significance.  However, it is also restricted to a kind of leading
approximation; they obtain a result for the DUPDFs from an examination
of the last step of the evolution.  This corresponds to one of our
extra factorization theorems for parton correlation functions, in its
leading-order approximation when the parton kinematics are strongly
ordered.  Our aims are more ambitious even if we have not achieved
them in QCD so far: we have obtained a formalism that can be applied
to arbitrarily non-leading corrections and in the context of a MCEG.
To this end, we have provided exact operator definitions of our
distributions.  In QCD, as we will explain, it is necessary to deal
with the issues symptomized by the gauge invariance problem of the
definition of the pCfs: Where should the Wilson lines be?  In a
leading logarithm approximation, these issues are readily evaded by
inserting appropriate cutoffs, without examining exactly how these are
to be implemented in an exact operator definition.  

The organization of the paper is the following. After explaining the
conceptual issues that concern us in Sec.\ \ref{sec:rationale}, we
describe our model theory in Sec.\ \ref{sec:dis}.  Then in Sec.\ 
\ref{sec:HS.factn} we derive our new factorization for DIS structure
functions in terms of parton correlation functions.  In Sec.\ 
\ref{sec:pcf}, we obtain factorization theorems for parton correlation
functions; these give both the showering algorithm and the computation
of parton correlation functions in terms of ordinary parton densities
(as in the \MSbar{} scheme).  In Sec.\ \ref{sec:imp} we show an 
algorithm for a MCEG that implements our factorization
formulas.  Then in Sec.\ \ref{sec:eg} we
give explicit calculations of the next-to-leading order (NLO) hard
scattering for deep-inelastic scattering.  Finally, in Sec.\ 
\ref{sec:cl}, we give our conclusions and an indication of the
approach we propose to use to extend our results to full QCD.

\FIGURE{
   \centering
   \psfrag{p}{$p$}
   \psfrag{q}{$q$}
   \psfrag{k}{$k$}
   \psfrag{k+q}{$k+q$}
% Force full width figure
\hspace*{1in}
   \includegraphics[scale=.8]{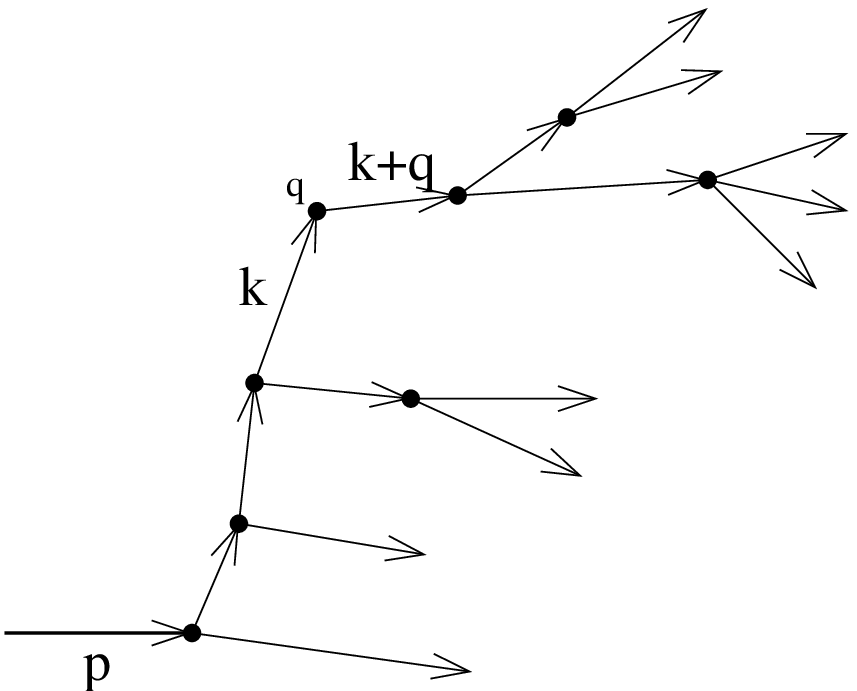}
\hspace*{1in}
    \caption{Generation and structure of event from MCEG for deep-inelastic
      scattering.  
      %The vertices represent the generation of a partonic
      %configuration, either a hard-scattering, labeled by $q$, or a
      %branching.  The lines represent partons or other particles.  
      One
      vertex has four attached lines; it would arise from a possible
      NLO branching.  
   \label{fig:event}
 } 
} 

%==========================================================
\section{Rationale for a new algorithm}
\label{sec:rationale}

The overall structure of a MCEG is that each event is generated by the
following steps:
\begin{enumerate}
\item Generate a hard scattering. This results in a set of partons
  and values of their momenta (or of some scalar variables that can be
  used to determine the momenta).
\item For each available parton, either choose not to shower it, or
  generate a branching, i.e., a set of new partons.
\item Go back to the previous step, applying it to any new partons
  that have been generated.
\end{enumerate}
This general structure corresponds to a factorization theorem, and the
generated events can be given a Feynman-graph-like structure, as in
Fig.\ \ref{fig:event} for deep-inelastic scattering (DIS).  The
generation of the hard scattering is represented by the vertex labeled
$q$, and the resulting partons are represented by the attached lines.
The branchings of these partons are represented by the vertices at the
other ends of the lines.  The arrows label momentum flow with respect
to a corresponding Feynman graph vertex.  The lines with no vertex at
the outgoing end are final-state particles, and the one line, labeled
$P$, with no vertex at the incoming end is the target particle.  Some
generalization is needed in the case of angular ordered coherent
branching, but this does not affect the overall ideas.  Notice that
the lines represent both the generated partons and the flow of
information between different units of computation, at the vertices.

\FIGURE{ \centering
  \psfrag{q}{$q^\mu$} \psfrag{p}{$P^\mu$} \psfrag{x}{$p_f^\mu$}
  \includegraphics[scale=.45]{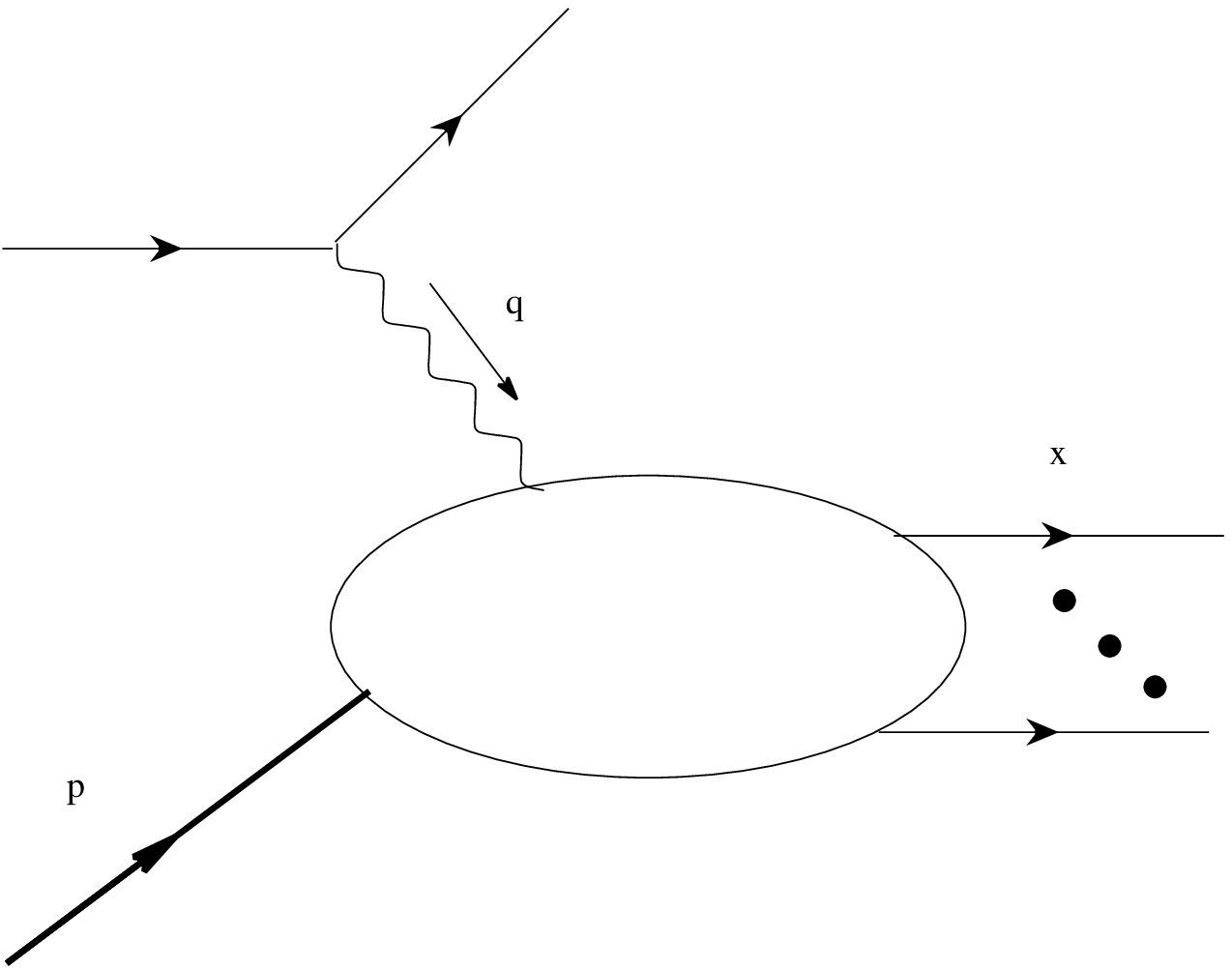}
\caption{Deep inelastic scattering.}
\label{fig:dis1}
}
The key property that gives a straightforward algorithm is that at
each step the showering probability for generating partons depends on
a fixed number of variables that correspond to the single parent
parton of that step.  The showering probability is, of course, a
conditional probability.\footnote{In a subtractive formalism beyond
  leading order, such as we will use, correction terms to the cross
  section may sometimes be negative, so we will allow the generation
  of weighted events, some with negative weights.  For this purpose,
  we must generalize the use of the term ``probability''.  Instead of
  a conditional probability, we will need what we might term a
  conditional weight, normalized like a probability so that its
  integral is unity.  }  Since at each step we have several partons,
each of which needs to be showered, the showering probabilities for
the different partons should be independent; this corresponds to a
factorization theorem.  What is needed to allow a clear systematic
discussion of non-leading corrections is a definite operator matrix
element expression for the probability for showering; any non-trivial
communication between the showers, as exhibited by a
non-factorization, will tend to preclude the possibility of a matrix
element formulation.  Without a systematic factorization,\footnote{%
  Of course, factorization does not hold as an exact statement.  It is
  only true as an approximation for certain important classes of
  kinematic region.  Also, when higher-order corrections are included,
  we obtain not an actually factorizing formula for the cross section
  but a sum of factorized terms, both for the hard scattering and for
  the parton branching.  } so that different showers are independent,
it is hard to give a systematic treatment for the production of
arbitrarily many particles.

As we will now show, the most obvious formulation for the algorithm,
founded on the parton model approximation to factorization, actually
violates some of these properties because of its treatment of parton
kinematics.  This is what motivates us to find a more general
approach.

For ease of presentation, we shall work with a model theory $\phi^3$
in space-time dimension $n=6$ throughout the paper, the methods 
immediately generalize to any nongauge theory. The details of the model
will be described in Sec.\ \ref{sec:dis}.

We consider DIS, whose kinematics are shown in Fig.\ \ref{fig:dis1}.
There $q$ is the photon momentum and $P$ the hadron momentum.  
We work in the virtual-photon--hadron
center-of-mass frame, using light-cone coordinates, $p^{\pm} = (p^0 \pm
p^z)/ \sqrt{2}$, where, up to corrections power-suppressed at large
$Q$,
\begin{eqnarray}
        q^\mu
        &=&
        \left( -xP^+, \frac{Q^2}{2xP^+}, \T{0} \right),
\\
        P^\mu
        &=&
        \left( P^+, \frac{M^2}{2P^+}, \T{0} \right),
\\
        x&\equiv&\frac{Q^2}{2P\cdot q} \,\, \mbox{and} \, \, 
        Q^2 \equiv -q^2.
\end{eqnarray}
\FIGURE{
  \centering
  \psfrag{q}{$q$}
  \psfrag{p_1}{$k$}
  \psfrag{p_{1'}}{\hspace*{-3mm}$k+q$}
  \psfrag{pp}{$P$}
  \psfrag{U}{D}
  \psfrag{L}{$\Phi$}

  \includegraphics[scale=0.5]{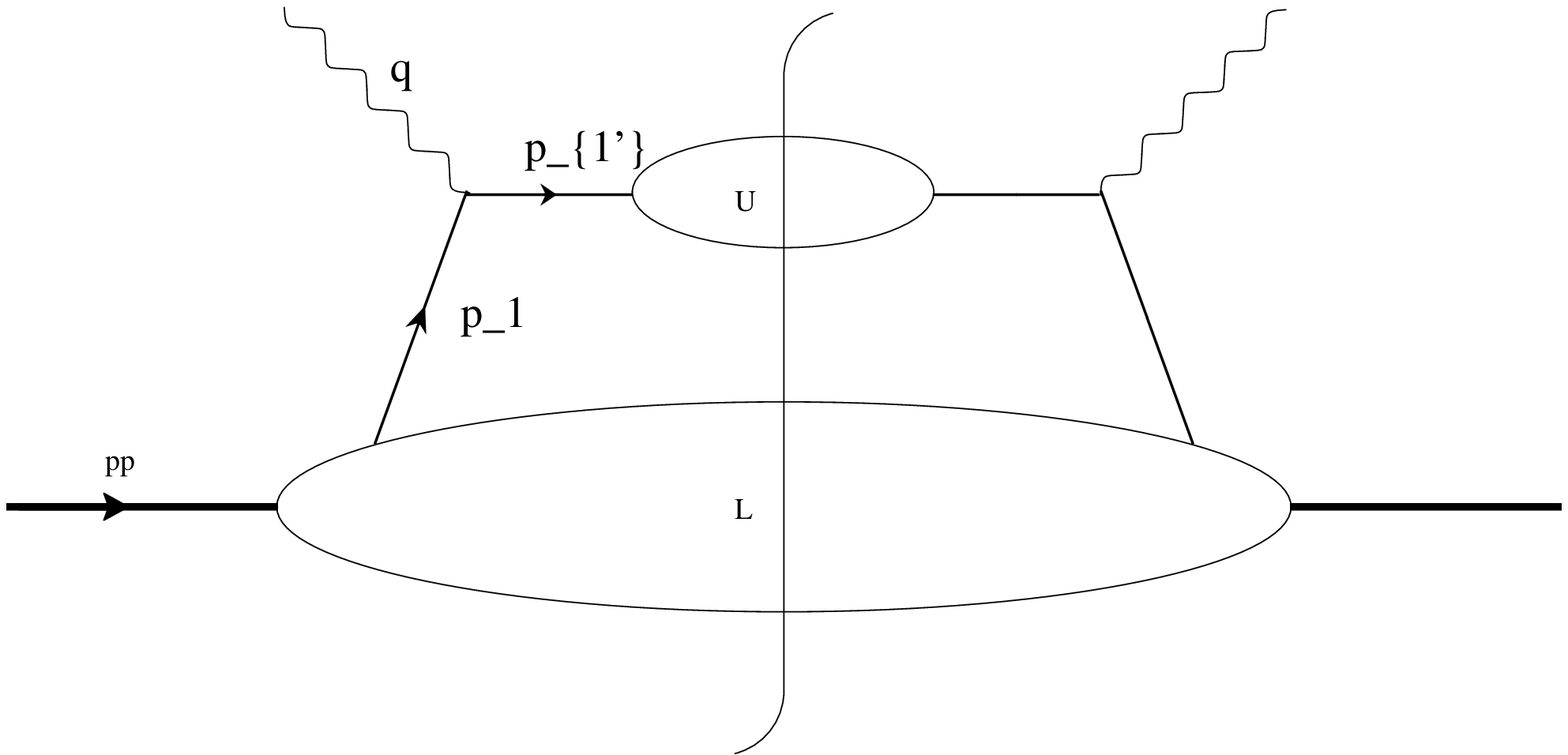}
  \caption{``Handbag'' diagram for DIS.}
  \label{fig:handbag}
}

The parton-model approximation is obtained from handbag diagrams, as
in Fig.\ \ref{fig:handbag}.  Notice that we have included a bubble
representing the full showering and hadronization of the struck quark.
Up to a normalization factor, which will not concern us here, the
value of the handbag is
\begin{equation}
\label{eq:parton.model.graph}
   F_2 = \frac{Q^2}{2\pi} \int \frac{ d^6k }{ (2\pi)^6 } \,
       \Phi(k, P) \, D(k) ,
\end{equation}
where $\Phi$ and $D$ correspond to sums over the correspondingly labeled
subgraphs in Fig.\ \ref{fig:handbag}, with an integral over all final
states.  The parton-model approximation \cite{LP} applies when $k$ has
low transverse momentum and virtuality compared with $Q$ and when the
struck quark also has virtuality much less than $Q$.  Then in the hard
scattering, which in this case is a trivial factor of unity, we can
replace the incoming and outgoing quarks by on-shell momenta 
\begin{equation}
  \label{eq:PM.momenta}
  \begin{split}
    k   &\simeq (xP^+,0,\T{0} ),
  \\
    k+q &\simeq (0,q^-,\T{0} ) = (0,Q^2/(2xP^+),\T{0} ).
  \end{split}
\end{equation}
However, this approximation cannot be made in the two nonperturbative
blobs $\Phi$ and $D$, but rather the small components of momentum must
be integrated over.  This leads to
\begin{eqnarray}
\label{eq:parton.model.formula}
   F_2 &\simeq& \left [ 
            xP^+ \int \frac{ dk^- \, d^4\T{k} }{ (2\pi)^6 }
            \Phi( xP^+,k^-,\T{k}, P )
       \right]
       \left[ \frac{2q^-}{2\pi} \int dl^+ D( l^+,q^-,\T{0} ) \right]
\end{eqnarray}
Up to a normalization, the first factor in square brackets corresponds
to the usual definition of a parton density, $f(x)$, as the matrix
element of a light-front operator.  The second factor is the integral
of the fragmentation function $D$ over final states, again up to a
normalization; this integral is over the discontinuity of a
propagator, so that the factor is equal to unity if the integral is
convergent.\footnote{There are, in fact, ultra-violet convergence
  problems in these approximated integrals, associated with the need
  for NLO corrections to the parton model; the divergences can be
  dealt with either by a cutoff or by renormalization \cite{factn}.
  But this issue will not affect this part of our discussion.  }

Notice that in the $\Phi$ and $D$ factors in the parton-model
approximation we have replaced the large components $k^+$ and
$k^-+q^-$ of the external momenta by the same approximate values
$xP^+$ and $q^-$ that are given in Eq.\ (\ref{eq:PM.momenta}); it is
only the small components that we leave unchanged.\footnote{We have
  also replaced the parton transverse momentum in the fragmentation by
  zero.  But this is simply equivalent to a small Lorentz
  transformation. 
}  We have therefore
replaced the final states of $\Phi$ and $D$ by final states with
different momenta.  For the inclusive structure function $F_2$, this is
satisfactory, since we are only concerned with the numerical values of
the factors in Eq.\ (\ref{eq:parton.model.formula}), not with the
detailed structure of the final state.  For this case, errors in the
approximation at large parton virtuality are properly compensated by
the usual correctly derived higher-order corrections to the hard scattering.

But the situation is different in a MCEG, as we will now see.  In the
first step of a MCEG, a hard-scattering configuration is generated
with a corresponding value of parton momentum $k$.  Since the
showering has not yet been performed, it is natural and conventional
to assign to this momentum the naive parton model value, as in Eq.\ 
(\ref{eq:PM.momenta}).  The cross section is expressed in terms of the
parton density $f(x)$, which, as in the first factor of Eq.\ 
(\ref{eq:parton.model.formula}), is defined in terms of an integral
over parton momentum $k$ with the plus component fixed: $k^+=xP^+$.
However the true value is different:
\begin{equation}
  \label{eq:true.k}
  k^+ = xP^+ + \frac{ M^2 + k_T^2 }{ 2 ( k^- + q^- ) },
\end{equation}
where $M$ is the invariant mass of the outgoing parton.  

Thus there is a mismatch between the true value and the initially
generated value of $k^+$ that is used to define the parton density.
This can be a large mismatch, since the virtualities and transverse
momentum range up to order $Q$, at least. The parton-model
approximation replaces the final state in blob $\Phi$ by a final state
with a different momentum.

Since in a MCEG we resolve the structure of the final state, the value
of $k$ has to be reassigned after showering to correspond to the true
final state.  The use of the word ``reassigned'' in the previous
sentence unmasks a conceptual shift that is critical in correctly
analyzing the algorithm in a MCEG.  There are in fact (at least) three
very different objects to which we can give the name $k$.  Two
correspond to the dummy variables of integration in one or other of
the two formulas (\ref{eq:parton.model.graph}) and
(\ref{eq:parton.model.formula}) for the handbag diagram; within a
Monte-Carlo implementation each of these variables can be treated as a
random variable which has a single definite (Lorentz-vector) value in each
event. The third object called $k$ is a storage location in the
computer program; its value can change during the generation of a single
event.  Such a reassignment of the computer variable changes the
correspondence between the computer variable and the random
variable(s) used to define probabilities.

Branching probabilities in initial-state showering
are conditional probabilities given the value of $k$, and in
particular given the value of $k^+$.  But the value of $k^+$ used in the
initial-state showering changes during showering, in a way that depends
on the shower history of the event. 
The same applies for the value of $k^-+q^-$ in the final-state 
shower. This makes the fundamental basis
of a conventional MCEG and of its independent showering probabilities
hard to quantify accurately enough to allow a transparent derivation
of higher-order corrections.

A useful characterization of the difficulty associates it with the
momentum-conservation delta function at the photon vertex.  Let $l$ be
the momentum of the outgoing quark.  Then the standard parton-model
approximation replaces
\begin{equation}
  \delta^{(6)}(k+q-l)
\end{equation}
by
\begin{equation}
  \delta(k^++q^+) \, \delta(q^--l^-) \, \delta^{(4)}(\T{l}).
\end{equation}
These are good approximations to each other only if they are
integrated with a sufficiently slowly varying function, as is the case
for the inclusive cross section.  But when we resolve the full final
state, the two delta functions are not good approximations to each
other; they are infinitely different.

Therefore we propose that the fundamental factorization formula
for a MCEG should involve unintegrated parton correlation functions,
and that showering then involve probabilities conditional on $k$ and
$k+q$ which have already been assigned their exact values.  While the
approximated momenta will still make their appearance, it will be in a
different and controllable fashion.

By using a factorization formula which does not contain ordinary parton
densities we have also removed the DGLAP evolution kernels, which
appear to form an essential part of the showering algorithm.  Other
quantities will take over their role.

We will find it convenient not to generate the exact parton momenta at
one step, but to reformulate in several substeps the first step of the
algorithm that we gave at the beginning of this section.  Thus we can
generate the hard scattering as follows:
\begin{itemize}
\item [1.a] Generate the hard scattering with conventional massless
  external partons, and conventional parton densities.\footnote{The
    parton densities should be the most accurate known parton
    densities, rather than just the approximation with LO evolution.}
\item [1.b] Generate, independently, the virtualities and transverse
  momenta of the partons according to some suitable approximate
  distributions.  
\item [1.c] Reassign the parton momenta to their exact values
  according to a defined prescription.
\item [1.d] Reweight (or appropriately veto) the event so that the
  probability/weight corresponds to the correct distribution in terms
  of parton correlation functions.  The values of the parton
  correlation functions will have been tabulated as the result of a
  separate calculation.
\end{itemize}
A similar procedure is applied at each step of parton showering.  The
motivation of this procedure is that the factorization formula in
terms of unintegrated parton correlation functions involves more
variables and more complicated objects than standard factorization.
So it is simpler to base the event generation on ordinary parton
densities and then correct the result to correspond to the new
factorization formula.

Although the reassignment of parton momenta and the reweighting are
superficially similar to those in other proposals, the important point
is that we propose a change in the logic.  Previously the hard scattering
and the branchings steps in showering generated approximate parton
momenta.  The reassignment of parton momenta to their exact values
occurred \emph{after} showering of these partons.  If we then try to
use a reweighting method to treat higher-order corrections for both
the hard scattering and for showering probabilities, there is overlap
between the parts of the event to which different reweightings are
applied.

We instead do the reassignment of parton momenta \emph{before} the
showering.  All the probability distributions for the actual parton
momenta are distributions as functions of the exact momenta.
Therefore the branching probabilities for different partons can be
taken as strictly independent of each other.  Moreover the reweighting
is not to make the differential cross section correspond to the
standard NLO calculation of the cross section but rather to make it
correspond to our new factorization formula in terms of parton
correlation functions.  Of course, the new formula will include the
information in the previous calculations, but it can potentially
include even higher order corrections.  Most importantly, the
application of reweighting to the hard scattering or one branching
does not overlap logically with later branchings.  Thus exactly similar
methods can be applied to both the hard scattering and to parton
branching, so that both can be systematically treated to arbitrary
order.

Our procedure is a generalization of the work by one of us in
\cite{phi3} for final-state showering.  The new feature is in a
certain sense concerned with the directions of cause and effect.  In
$e^+e^-$-annihilation we generate a hard scattering given the incoming
leptons.  We shower the partons and readjust their kinematics by some
chosen prescription to satisfy momentum conservation.  

But with an initial-state hadron, this procedure requires us to
readjust the kinematics of the \emph{incoming} parton whose value of
$k^+$ was used to determine the cross section from the parton
density.  The transformation of the kinematics to get correct parton
momenta inevitably changes the value of $k^+$ --- see Eq.\
(\ref{eq:true.k}).  This invalidates the value of the non-perturbative
parton density used at the initial step of computing the
hard-scattering probability.  The corresponding non-perturbative
probability in $e^+e^-$-annihilation is the unit probability for a
parton to shower.

%==========================================================
\section{The model for DIS}

\label{sec:dis}
Our method in its present form works for any non-gauge theory.  But
for simplicity of presentation, we work with $\phi^3$ theory in 6
space-time dimensions.  The Lagrangian is
\begin{eqnarray}
\label{eq:lagr}
    \mathcal{L}
&=&
    \frac{1}{2}(\partial \phi)^2 
    - \frac{1}{2}m^2\phi^2
    -\frac{1}{6} g\left( \frac{\mu^2e^\gamma}{4\pi} \right)^{\epsilon/2}
     \phi^3
    + \mathcal{L}_{ct},
\nonumber \\
    \mathcal{L}_{ct}
&=&
    \frac{1}{2} \delta Z_2(\partial \phi)^2 
    - \frac{1}{2} \delta m^2\phi^2
    - \frac{1}{6} \delta g\phi^3 
    - \delta h\phi .
\end{eqnarray}
For regularization of ultra-violet divergences we use a space-time
dimension $n=6-2\epsilon$. We associate a factor
$\bar{\mu}^{\epsilon}\equiv[\mu^2e^\gamma/(4\pi)]^{\epsilon/2}$ with the coupling so that
\MSbar{} renormalization is implemented by pure pole counterterms.  We
use \MSbar{} renormalization for all except the renormalization of
tadpole graphs, and for those we define the $\delta h\phi$ counterterm
\cite{jccbook} by requiring the sum of the tadpoles and their 
counterterms to vanish, i.e., by the renormalization condition 
$\langle0|\phi|0\rangle = 0$.  
We define a bare parton field as $\phi_0=Z^{1/2}\phi$. 

Our model for DIS consists of the exchange of a weakly interacting
boson, $A$, for which we add an extra term to the interaction
$\mathcal{L}_{s}= e_0 A \phi_0^2/2$ and a corresponding counterterm.  As
in \cite{phi3}, we unify the calculation of all cross sections in a
weighted cross section
\begin{eqnarray}
     \sigma[W]
&=&
      K \sum_f \langle P|j(0)|f\rangle \langle f|j(0)|P\rangle \,\,
      (2\pi)^6 \delta^{(6)}(q+P-p_f) \,\, W(f) 
\nonumber \\
&=&
      K \sum_f W(f)\int d^6 y\,\, e^{iq\cdot y}\,\,
      \langle P| j(y)|f\rangle \langle f| j(0) |P \rangle,
\end{eqnarray}
where the weighting function $W(f)$ is an arbitrary smooth function of
the final state.  The current
$j=e\frac{1}{2}[\phi^2]=e_0\frac{1}{2}\phi_0^2$ contains a renormalized
composite operator \cite{jccbook} $[\phi^2]$, and $K$ is a standard
leptonic factor.  We have chosen to include a factor of the analog of
the electromagnetic coupling in the operator.  This ensures that as
regards strong interactions the operator is scale invariant; the
``electromagnetic'' coupling $e$ in our model gets renormalized by
strong interaction effects unlike the case of the true electromagnetic
coupling in QCD. 

Any exclusive component of the cross section can be obtained from
$\sigma[W]$ by functional differentiation with respect to $W$, while,
by the choice of a suitable form for $W(f)$, jet cross sections and in
general any kind of binned cross section, can be obtained.  The use of
weighted cross sections is very compatible with the MCEG approach,
since the output of a generator is a list of events with a
distribution corresponding to an approximation to the fully
differential cross section.  An estimate of $\sigma[W]$ is obtained
from a sum over the generated events weighted by $W(f)$:
\begin{equation}
\label{sec:sigma.MC}
   \sigma_{\rm est}[W] = \frac{1}{L} \sum_i W(f_i),
\end{equation}
where $f_i$ is the final-state of event $i$, and $L$ is the luminosity 
appropriate for the set of events.  If the EG gives weighted events,
then a factor of the event weight $w_i$ needs to be inserted inside
the summation.

%==========================================================
\section{Hard-scattering factorization} 
\label{sec:HS.factn}

The key idea of our method is that all partons should be generated
with their exact kinematics.  So in this section we will obtain a
factorization theorem, Eq.\ (\ref{eq:fact1a}) below, which expresses
the general weighted cross section $\sigma[W]$ in terms of parton
correlation functions.  

While we will use the new factorization property for generating the partonic
content of DIS events, we will also need the usual factorization
theorem to determine values of the inclusive cross section:
\begin{equation}
  \label{eq:fact.incl}
  \sigma[1] = K \int_x^1 \frac{d\xi}{\xi} \, 
         C\boldsymbol( x/\xi,Q^2/\mu^2,g(\mu) \boldsymbol) \, f(\xi;\mu) 
  + \mbox{power-suppressed correction}
  ,
\end{equation}
where $C$ is the hard scattering coefficient and $f$ is the ordinary
parton density in the target.  We choose to define the parton density
in the \MSbar{} scheme.\footnote{As usual, the dependence of $f$ on the
  renormalization scale $\mu$ includes the effect of the running
  coupling and mass $g(\mu)$ and $m(\mu)$, and is governed by a
  DGLAP equation.}  The coefficient $C$ is perturbatively calculable
if $\mu$ is chosen to be of order $Q$.  Notice that this theorem only
applies after an unweighted sum over final states, so the argument of
$\sigma$ is a weight function whose value is unity for all states.

The normalizations are chosen so that the lowest order value of the
hard-scattering coefficient, associated with the handbag diagram with
neglect of hadronization, reproduces the parton model:
\begin{equation}
  C(x/\xi,Q^2/\mu^2,g) = e^2 \xi \delta(\xi-x) + O(e^2g^2),
\end{equation}
so that
\begin{equation}
  \sigma[1] = K \, e^2 \, f(x;\mu)  \,\, 
         \left[ 1 + O\left(g^2(Q)\right) \right].
\end{equation}

%-----------------------------------------------
\subsection{Regions}

\FIGURE{
\centering
\psfrag{q}{ $q$}
\psfrag{p}{ $P$}
\psfrag{1}{\tiny $U_1$}
\psfrag{n}{\tiny $U_{N_u}$}
\psfrag{k}{ $k$}
\psfrag{h}{\scriptsize $H$}
\psfrag{e}{$=$}
\psfrag{sum}{}
\psfrag{u}{ $U(q,k)$}
\psfrag{f1}{$F(k,P)$}
   \VCbox{\includegraphics[scale=.5]{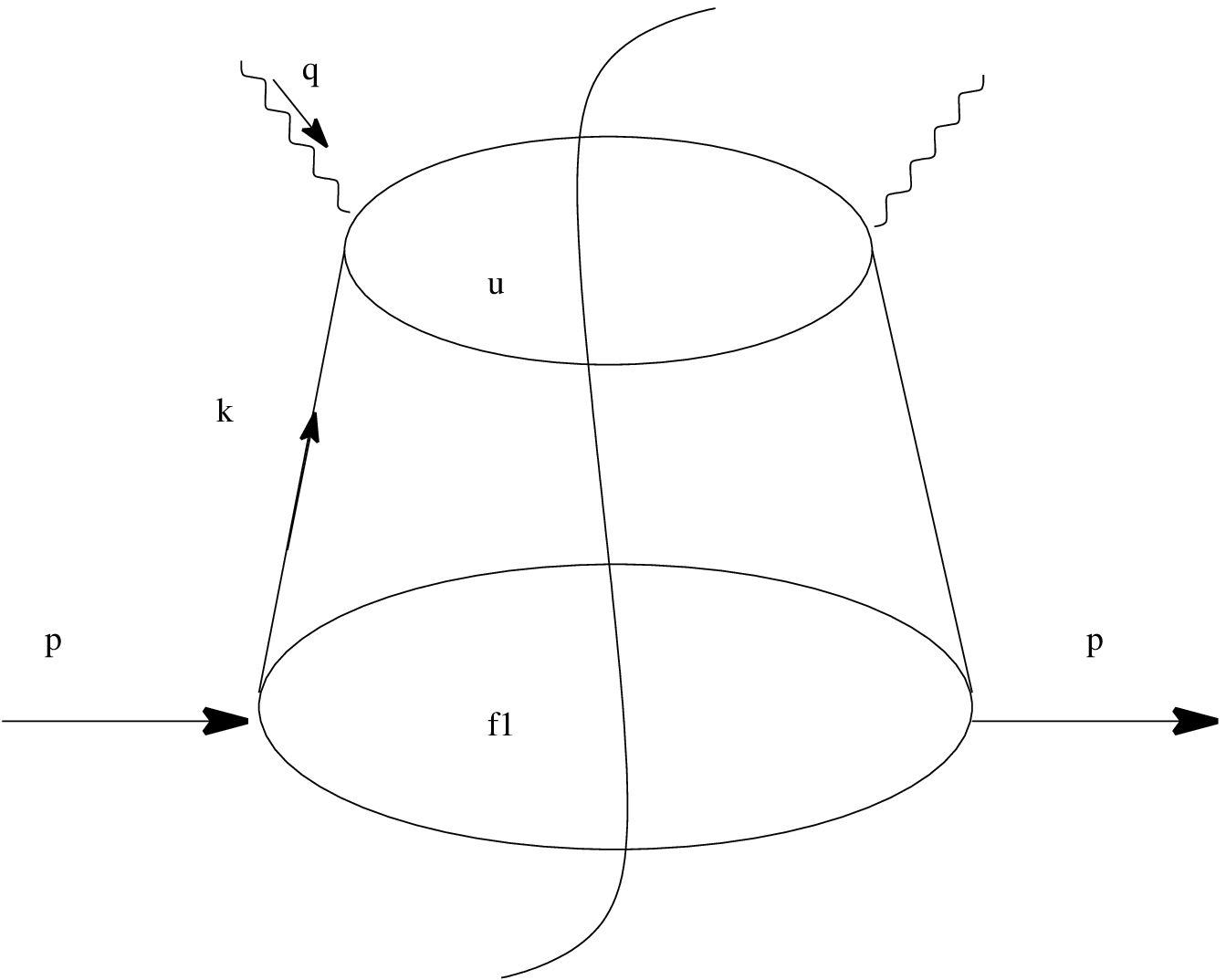}}
=
   \VCbox{\includegraphics[scale=.5]{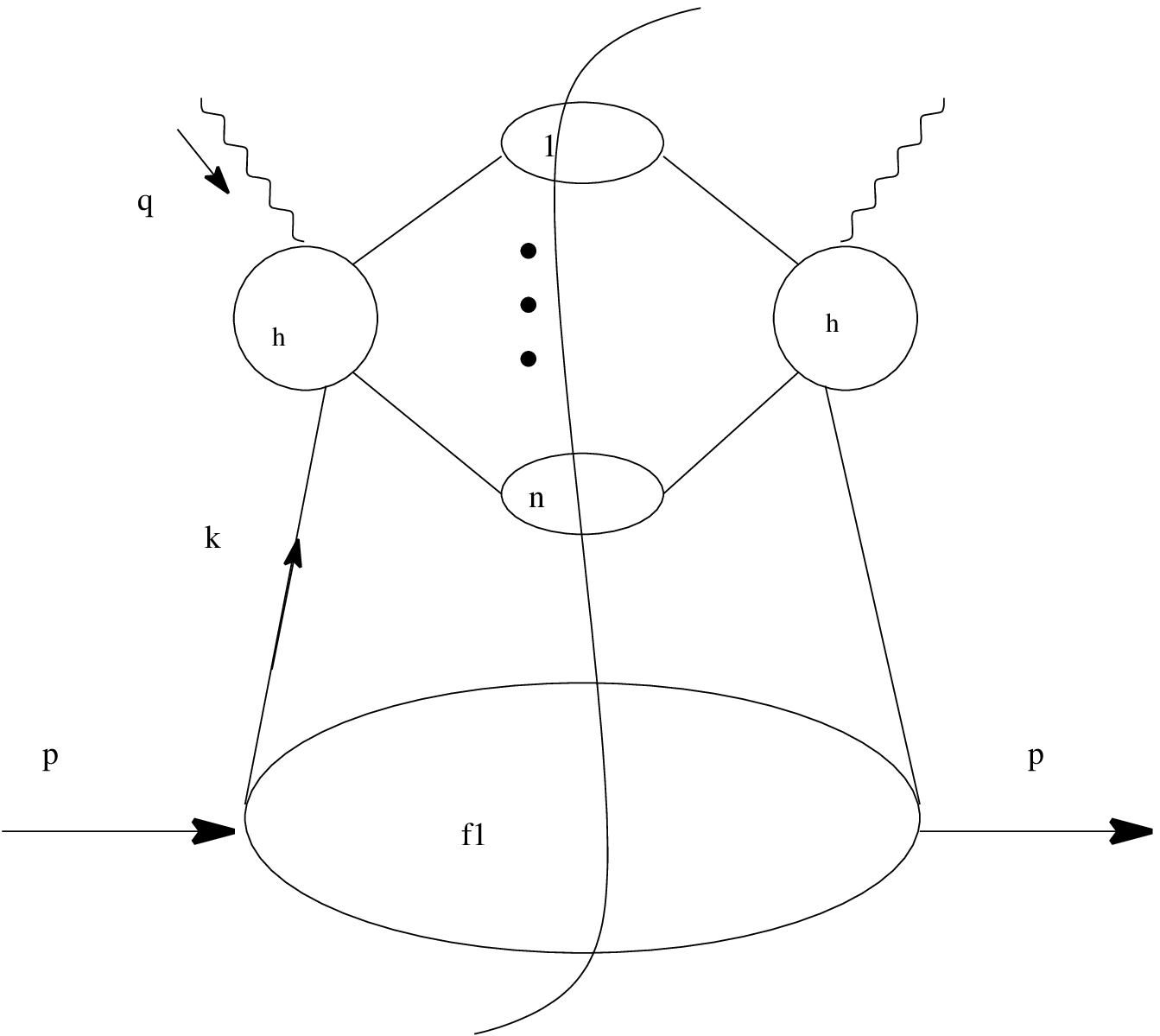}}
  \caption{Decomposition of a graph for DIS corresponding to a
    generic leading region.  
  \label{fig:dis2}
  }
}

\noindent
The regions that give leading power contributions for a DIS process
are obtained by standard arguments \cite{sterman} and are illustrated
in Fig.\ \ref{fig:dis2}: Momenta in the lower subgraph $F$ are
approximately in the proton direction, and momenta in the upper
subgraph $U$ are grouped around one or more different
directions.  The hard subgraph $H$ has internal momenta that are all
off-shell by order $Q^2$.  If momenta could be cleanly separated into
different regions, a sum over all graphs and over all regions for each
graph would give quite directly the factorization formula, Eq.\ 
(\ref{eq:fact1a}) below, which corresponds to the factorized structure
in Fig.\ \ref{fig:dis2}.  Moreover interpreting the blobs in 
Fig.\ \ref{fig:dis2} as complete amplitudes, not just as sums over Feynman
graphs, strongly indicates what factorization should mean
non-perturbatively in the full theory, even if we are not yet able to
make a real proof beyond perturbation theory.

However, there are momentum configurations that interpolate between
core parts of the different regions and these give well-known
logarithmic enhancements, evolution of parton densities, etc.  So an
unambiguous and clean separation between different regions of momentum
does not exist.  All but the simplest graphs for the cross section (or
structure function) have several leading regions.  We will analyze
them region by region, obtaining an expression for each graph $\Gamma$
as a sum of terms, one for each region $R$:
\begin{equation}
  \label{eq:regions}
  \Gamma = \sum_R C_R(\Gamma) + \mbox{power-suppressed correction},
\end{equation}
with the application of subtractions to compensate double counting
between regions.  The principles of the argument will be exactly those 
of our previous work \cite{phi3} for $e^+e^-$ annihilation.  The
changes will be those to accommodate an initial-state hadron and an
algorithm for backward showering for the initial state.

We will find that subtractions only need to be applied inside the hard
scattering, so that summing over graphs and regions will give a
factorized form for the weighted cross section.  Schematically
\begin{equation}
\label{eq:fact.HS}
       \sigma[W] = K \sum_f 
       \sum_{\substack{\text{graphs }\Gamma\\\text{regions }R}} 
          \hat{H}_R 
         \prod_{j=1}^{N_u} U_{Rj} \, F_R \, W(f) + 
        \mbox{power-suppressed correction}.
\end{equation}
We use a subscript $R$ to denote the values of quantities that are
appropriate to a particular region of a particular graph.  This will
allow us to use the unsubscripted values for the corresponding
quantities $H$, and $F$, etc, summed over the relevant subgraphs, that
appear in the final factorization formula.  
Thus $\hat{H}_R$ is contribution to the subtracted hard scattering
factor, from the 
$H$ subgraph in Fig.\ \ref{fig:dis2}.  Similarly $F_R$ and $U_{Rj}$
correspond to the target and final-state ``jet'' factors.  We have
made explicit the sum/integral over final states, so that we can
insert a non-factorized weight $W(f)$ to analyze the final-state
structure.  We will see that the sum over graphs can be performed
separately for $\hat{H}_R$, $F_R$ and $U_{Rj}$, and this will lead to the
actual factorization formula, Eq.\ (\ref{eq:fact1a}) below.

%----------------------------------------------------------
\subsection{Factorized approximation for one region}
\label{sec:def}
We now construct a factorized approximation for a general leading
region $R$ of a general graph $\Gamma$ for DIS.  The approximation
must be compatible with the structure of a MCEG, as described in 
Sec.\ \ref{sec:rationale}, which means that it must have a factorized form
and must preserve exact parton momenta for the final state.  The
accuracy of the approximation will degrade as momenta move away from
the core of the region, and the full factorization theorem will
contain subtractions in the definition of the hard scattering
coefficient that eliminate double counting when we combine the
contributions of different regions.  Then the final answer will be
uniformly accurate in all the different regions and in the
interpolating regions.

It is useful to define a systematic notation for the momenta in Fig.\
\ref{fig:dis2}: 
\begin{itemize}
   \item $N_u$ is the number of the final-state partons
          connecting the hard scattering and the 
          upper ``jet'' subgraphs in 
          $U(q,\,k,\,m)$.
   \item $n_j$ is the number of final-state particles arising from parton
          $j$ in subgraph $U_j$.
   \item The momenta of these particles are
              $p_{j,1}$, $p_{j,2}$, \ldots, $p_{j,n_j}$.
   \item $\underline{p}_j$, $j=1, \ldots , N_u$, with an
              underline, denotes the {\it collection} of the
              final-state particle momenta in the jet subgraph $U_j$, i.e.,
              $\underline{p}_j = ( p_{j,1},\,p_{j,2}, \ldots,\, p_{j,n_j})$.
   \item $l_j \equiv \sum_{i=1}^{n_j} p_{j,i}$ denotes the momentum 
        of parton $j$, and $M_j$ denotes its invariant mass.
   \item $\underline{l}_u = (l_1,\,l_2, \ldots,\, l_{N_u})$ denotes the
        {\it collection} of the final-state parton momenta. 
   \item $\underline{M} = (M_1, \, M_2, \, \ldots, \, M_{N_u})$ denotes the
        {\it collection} of the final-state parton invariant masses.
   \item $\underline{\underline{p}}_u=(\underline{p}_1,\,
        \underline{p}_2, \ldots, \, \underline{p}_{N_u}) $ denotes the
        \emph{collection} of \emph{all} the final-state particles in the
        upper part of the graph, $U(q,k,m)$.
   \item Similarly we use index $j=0$ for the target subgraph $F$, {\it i.e.}
        \begin{itemize}
        \item $n_0$ is the number of final-state particles in the target 
            subgraph $F$.
        \item $\underline{p}_0$ denotes the {\it collection} of the final-state
        particle momenta in $F$.
        \end{itemize}
\end{itemize}
We introduce the following notations for the final-state phase space:
\begin{eqnarray}
    dL\!\left( \underline{\underline{p}}_u;q+k;m_{\rm ph} \right)
&=&
    \prod_{j=1}^{N_u} \prod_{i=1}^{n_j} 
         \frac{ d^5\vec{p}_{j,i} }
              { (2\pi)^5 2 \sqrt{\vec{p}_{j,i}^2+m_{\rm ph}^2} }
    \,\,
    (2\pi)^6 \delta^{(6)}( {\textstyle\sum_{j,i}} p_{j,i} - q -k ) ,
\nonumber \\
    dL( \underline{p}_j;l_j;m_{\rm ph} )
&=&
    \prod_{i=1}^{n_j} 
        \frac{ d^5\vec{p}_{j,i} }
             { (2\pi)^5 2 \sqrt{\vec{p}_{j,i}^2+m_{\rm ph}^2} }
    \,\,
    (2\pi)^6 \delta^{(6)}( {\textstyle\sum_i} p_{j,i} - l_j ) ,
\nonumber \\
    dL\!\left( \underline{l}_u;q+k;\underline{M} \right)
&=&
    \prod_{j=1}^{N_u} 
        \frac{d^5\vec{l}_j}{(2\pi)^5 2 \sqrt{\vec{l}_j^2+M_j^2}}
    \,\,
    (2\pi)^6 \delta^{(6)}( {\textstyle\sum_j} l_j - q - k ) .
\end{eqnarray}
Here  $m_{\rm ph}$ is
the physical mass of the true final-state particles.
As pointed out in \cite{phi3}, the integration over the final-state phase 
space is factorizable:
\begin{eqnarray}
        dL\!\left( \underline{\underline{p}}_u;q+k;m_{\rm ph} \right)
      &=&\prod_{j=1}^{N_u} \prod_{i=1}^{n_j} 
          \frac{d^5\vec{p}_{j,i}}{(2\pi)^5 2E_{j,i}} 
    \,\,
        (2\pi)^6 
        \delta^{(6)}( {\textstyle\sum_{j,i}} p_{j,i} - q -k )
\nonumber \\
        &=& \left( \prod_{j=1}^{N_u} \int \frac{dM_j^2}{2\pi} 
                   \int \frac{d^5\vec{l}_j}
                          {(2\pi)^5 2 \sqrt{\vec{l}_j^2+M_j^2}}
           \right)
    \,\,
           (2\pi)^6 \delta^{(6)}( {\textstyle\sum_j} l_{j} - q - k ) 
\nonumber \\
        &&\times \prod_{j=1}^{N_u} 
            \left[ 
                  \prod_{i=1}^{n_j} 
                       \frac{d^5\vec{p}_{j,i}}{(2\pi)^5 2 E_{j,i}} 
    \,\,
                  (2\pi)^6 \delta^{(6)}( {\textstyle\sum_i} p_{j,i} - l_j ) 
            \right]
\nonumber \\
        &=& \left( \prod_{j=1}^{N_u} \int \frac{dM_j^2}{2\pi}\right)
            \int dL\!\left( \underline{l}_u;q+k;\underline{M} \right)
            \prod_{j=1}^{N_u} dL\!\left( \underline{p}_j; l_j; m_{\rm ph} \right) .
\end{eqnarray}
With this notation, we now write a decomposition for a graph $\Gamma$
and a particular region $R$ of the form in Fig.\ \ref{fig:dis2}:
\begin{eqnarray}
\label{eq:Gamma.R}
        \Gamma[W]&=& \int \frac{d^6k}{(2\pi)^6} \,\,
        dU_R\!\left( \underline{\underline{p}}_u,k \right) \,\, 
        dF_R\!\left( \underline{p}_0, k,P \right) \,\, W(f)
\end{eqnarray}
with
\begin{align}
\label{eq:dU}
     dU_R\!\left( \underline{\underline{p}}_u, k \right) 
& = 
     dL\!\left( \underline{\underline{p}}_u;q+k;m_{\rm ph} \right)
     \prod_{j=1}^{N_u} U_{Rj}\!\left( \underline{p}_j \right) 
   \,\, H_R\!\left( \underline{l}_u,\,k;\,q;\,m \right) ,
\\
\label{eq:dF}
     dF_R\!\left( \underline{p}_0, k, P \right) 
& = 
    dL\!\left( \underline{p}_0; P-k; m_{\rm ph} \right) \,\,
    F_R\!\left( \underline{p}_0,P \right) .
\end{align}
Note that at this point we
have not changed or approximated the graph in any way; we have simply
written the graph in a form suitable for analyzing it in region $R$.

In \cite{phi3}, integrated jet factors were defined corresponding to
sums of each of $U_{Rj}$ over graphs and final states.  These serve as
normalization factors in the MC algorithm for showering.  Because of
the different organization of the algorithm in this paper, our
definitions will be slightly different.  First we define an
integrated jet factor as
\begin{equation}
\label{eq:JI.def}
        J_I( l^2; \mu ) 
= \int d^6y \,\, \langle 0 | \phi(y) \phi(0) |0 \rangle
        \,\,e^{il\cdot y} .
\end{equation}
For consistency with our naming for the target-related factor, we can
also call this a ``vacuum parton correlation function'' or a
``final-state parton correlation function''.

Similarly, for the target factor, we define a parton correlation
function (pCf) $\Phi(k,P)$ as the full non-perturbative quantity that
corresponds to integrating $F$ over all final states and then summing
over graphs:
\begin{eqnarray}
\label{eq:pcf}
   \Phi(k,P)
   &=& \sum_{\substack{\text{graphs}\\\text{final~states}}} 
    dL\!\left( \underline{p}_0; P-k; m_{\rm ph} \right) \,\,
    F_R\!\left( \underline{p}_0,P \right) 
\nonumber\\
   &=& \int d^6y \,\, \langle P| \phi(y) \phi(0) |P \rangle
        \,\,e^{-ik\cdot y} .  
\end{eqnarray}
The pCf is necessarily nonnegative:
\begin{eqnarray}
\label{eq:pcf0}
      \Phi(k,P) 
  &=& \sum_Y \int d^6y\,\, \langle P| \phi(0)|Y\rangle \,
         e^{i(P-P_Y)\cdot y} \, \langle Y|\phi(0) |P \rangle
        \,\,e^{-ik\cdot y} 
\nonumber \\
  &=& \sum_Y (2\pi)^6\delta(k+P_Y-P) \,
         \left| \langle P|\phi(0)|Y\rangle \right|^2 
\nonumber\\
  &\geq& 0,
\end{eqnarray}
as is the integrated jet factor $J_I$.

We could now formulate factorization with an off-shell hard
scattering.  But on-shell and massless amplitudes are easier to
compute.  So we will now define an approximation appropriate to region
$R$ in terms of a hard scattering defined with massless external 
partons.  We will denote the approximation by an operation $T_R$:
\begin{eqnarray}
\label{eq:approx}
       T_R \Gamma[W] &\equiv& \int \frac{d\hat{k}^+}{2\pi}
           \int \frac{ dk^- \, d^{4}\T{k} }{ (2\pi)^5 }
           \int dL(\underline{p}_0;P-k,m_{\rm ph}) \,\,
           F_R(\underline{p}_0,P) 
\nonumber \\
        &&\times \int dL(\hat{\underline{l}}_u;\hat{k}+q; M=0)  
        \,\, H_R(\hat{\underline{l}}_u;\hat{k}, m\to0)
\nonumber \\
        &&\times  
        \prod_j \left[ \int \frac{dM_j^2}{2\pi} 
                  \int dL(\underline{p}_j;l_j;m_{\rm ph}) \, \Theta(M_j^2/\mu_J^2) \,
                  U_{Rj}(\underline{p}_j)
           \right]
      \,\, W(f) .
\end{eqnarray}
In this formula we have used a projection of the true parton momenta
$k$, $l_1$, \ldots, $l_{N_u}$ onto massless momenta $\hat{k}$,
$\hat{l}_1$, \ldots, $\hat{l}_{N_u}$, with zero transverse momentum for
$k$; we will define such a projection in Sec.\ \ref{sec:proj}.  We
have defined the hard scattering to be computed with massless external
momenta and with zero mass parameter for the internal lines.  However,
the target factor $F_R$ and the final-state jet factors $U_{R_j}$ are
computed with the exact parton momentum $k$ and $\underline{l}_u$, 
not the approximation.

As in \cite{phi3}, there is a cut-off function $\Theta(M_j^2/\mu_J^2)$ that
restricts the final-state parton masses to be below some factorization
scale\footnote{Conventionally, the term ``factorization scale''
  applies to a quantity analogous to $\mu_J$ but for the treatment of
  initial-state parton radiation in the conventional factorization
  formalism.  But the same principles apply in fragmentation functions
  and the like in the final state, so we will use the same name.  In
  \cite{phi3}, the symbols $\mu_F$ and $\mu_R$ were used to denote the
  quantities that are $\mu_J$ and $\mu$ in this paper. }
$\mu_J$, which should be of order $Q$.   An ordinary $\theta$-function
$\Theta(M_j^2/\mu_J^2)=\theta(1-M_j^2/\mu_J^2)$ would be suitable, although a smooth
cut-off might be better for implementation in MCEGs. 
After we construct a factorization formula, this cutoff function
will also appear in the subtractions in the hard scattering.  Since
the cutoff function does not appear in the definition of the cross
section itself, there is a kind of generalized renormalization-group
invariance: Changes in the value of $\mu_J$ and in the functional form
of $\Theta$ will leave the factorized approximation to the cross section
unchanged, up to power-law corrections, when all orders of
perturbation theory are included.  

A similar cut off function could be applied to the initial state, 
but the coupling of the kinematics of initial and final state showers
(through energy-momentum conservation) combined with our use of 
exact parton kinematics from the beginning for the struck
parton's momentum $k$, removes the need to do so.

%Conventionally, the term ``factorization scale'' applies to a quantity
%analogous to $\mu_J$ but for the treatment of initial-state parton
%radiation in the conventional factorization formalism.  But the same
%principles apply in fragmentation functions and the like in the final
%state, so we will use the same name.

Changing integration variables to the exact parton momenta gives
\begin{eqnarray}
\label{eq:approx1}
       T_R \Gamma[W] &=& 
           \prod_j \int \frac{d^6l_j}{(2\pi)^6}
           \int dL(\underline{p}_0;P-k,m_{\rm ph}) \,\,
           F_R(\underline{p}_0,k,P) 
      \,\,  H_R(\hat{\underline{l}}_u;\hat{k}, m\to0)
        \,\, \Delta_{N_u}(\underline{l}_u,k)
\nonumber \\
        &&\times  
        \prod_j \left[ \int dL(\underline{p}_j;l_j;m_{\rm ph}) \Theta(M_j^2/\mu_J^2) 
                  U_{Rj}(\underline{p}_j)
           \right]
      \,\, W(f) ,
\end{eqnarray}
with $k=\sum_jl_j-q$.  The quantity $\Delta_{N_u}$ is the Jacobian of the
transformation between the exact momenta and the massless approximate
momenta --- see Eq.\ (\ref{eq:jac}) below for its value.

In a region where the final-state lines of the subgraphs $U_{Rj}$ do
in fact form jets and where $|k^2|,\T{k}^2 \ll Q^2 $, $T_R\Gamma$
provides a good approximation to the original graph $\Gamma$, up to
power-law corrections (and the Jacobian approaches unity).

Now when we treated $e^+e^-$ annihilation in Ref.\ \cite{phi3} we
found a convenient and simple factorized approximation in terms of the
massless momenta.  It allowed the MC generation of massless parton
configurations for the hard scattering, with probabilities independent
of the showering.  But in DIS, the coupling of the kinematics of final
and initial-state partons, as given in Sec.\ \ref{sec:rationale},
prevents us from having a correspondingly simple decoupling of
showering from the generation of massless partons for the hard
scattering.  So it is not so clearly advantageous which set of
variables to use here.  It follows that there is also a choice of
which formula for $T_R\Gamma$ is defined to contain the Jacobian 
$\Delta_{N_u}$.  Any
change here is compensated for in physical cross sections by the
subtractions in higher-order corrections to the hard scattering, so
this is primarily a matter of convenience; the Jacobian is unity in
the collinear region.  Our choice probably simplifies the logic of
deriving subtractions for hard scattering, since they most naturally
involve massless approximations at certain points.  See Sec.\
\ref{sec:schemes} below for further details.

%----------------------------------------------------------
\subsection{Definition of projection onto massless momenta and its Jacobian}
\label{sec:proj}
In this section we give one possible definition of the projection onto
massless parton momenta: $\hat{k}\equiv P_{N_u}(k)$,
$\hat{\underline{l}}_u \equiv P_{N_u}(\underline{l}_u)$.  We require it to
obey
\begin{align}
\label{eq:cons1}
    \hat{k}^- &= \hat{k}_T =0;
\\
    \hat{l}_j^2&=0, \,\,(j=1, \ldots, N_u).
\end{align}
We also require energy-momentum conservation, so that $\sum_j \hat{l}_j
= q+\hat{k}$.  This gives the following constraints;
\begin{eqnarray}
\label{eq:cons2}
        -k^- + \sum_j l_j^- &=& \sum_j \hat{l}_j^- = q^- = \frac{Q^2}{2xP^+};
        \nonumber \\
        -k_T + \sum_j l_{jT} &=& \sum_j \hat{l}_{jT} =\T{q}=0;
        \nonumber \\
        -k^+ + \sum_j l_j^+ &=& -\hat{k}^+ +\sum_j \hat{l}_j^+ = q^+ = -xP^+.
\end{eqnarray}
In addition, this projection should have a smooth limit when some or
any of the external momenta go to zero.  Any projection would be a
valid choice if it satisfies the above constraints and 
$P_{N_u}[P_{N_u}(l,\,k)]=P_{N_u}(l,\,k)$.  It is also desirable that
the formulas be invariant under boosts in the $z$ direction. 

Our chosen definition is:
\begin{align}
\label{eq:proj}
  & \begin{cases}
           \hat{k}_T =0, 
     \\
           \hat{l}_{jT}= l_{jT} - k_T/N_u,
     \end{cases}
\nonumber \\
   & \begin{cases}
         \hat{k}^- =0,
     \\
              \hat{l}_j^-= l_j^- - k^-/N_u,
     \end{cases}
\nonumber \\
   & \begin{cases}
        \displaystyle
        \hat{l}_j^+ = \frac{\hat{l}_{jT}^2}{2\hat{l}_j^-} = 
        \frac{(l_{jT}-k_T/N_u)^2}{2(l_j^--k^-/N_u)}, 
     \\
        \displaystyle
        \hat{k}^+ = \sum_{j=1}^{N_u} \hat{l}_j^+ -q^+
         =\sum_{j=1}^{N_u} \frac{(l_{jT}-k_T/N_u)^2}{2(l_j^--k^-/N_u)}
                   -q^+.
     \end{cases}
\end{align}
Given the values of $k^-$, $k_T$ and of the $M_j$s, which parameterize
the deviation of the true parton momenta from the projected momenta,
we find the inverse transformation:
\begin{align}
\label{eq:invtrans}
        l_{jT} &= \hat{l}_{jT} + k_T/N_u, 
     &
        l_j^- &= \hat{l}_j^- + k^-/N_u, 
\nonumber \\
        l_j^+ &= \frac{ l_{jT}^2 + M_j^2 }{ 2l_j^- },
     &
        k^+ &= \sum_j l_j^+ -q^+.  
\end{align}
The kinematics of DIS requires that $k^2 < 0$, so either a rejection
or a good selection method is needed to eliminate positive $k^2$
events.

Notice that, although DIS kinematics require that $q_T=0$, our choice of
projection applies to $q_T\neq0$ without any modification.
 
We need the Jacobian of the transformation between the exact and the
approximated parton momenta.  We obtain it by expressing
$dL(\underline{l})$ in terms of $-$ and $T$ components of momentum,
and then using the fact that at fixed $k^-$ and $\T{k}$, the
transformation from $l_j^-$ and $l_{jT}$ is a simple shift.  Thus
\begin{eqnarray}
       dk^+ dL(\underline{l}_u;q+k;\underline{M})
  &=&
       dk^+ \prod_{j=1}^{N_u-1} \int \frac{dl_j^- d^4l_{jT}}{(2\pi)^52l_j^-}
        \frac{1}{2l^-_{N_U}} 
        2\pi\, \delta\!\left( \sum_j l_j^+ -q^+ -k^+ \right)
\nonumber \\
  &=&
       dk^+
        \prod_{j=1}^{N_u-1} \int \frac{d\hat{l}_j^- d^4\hat{l}_{jT}}{(2\pi)^52l_j^-}
        \frac{1}{2l^-_{N_U}} 
        2\pi\, \delta\!\left( \sum_j l_j^+ -q^+ -k^+ \right)
\nonumber \\
  &=&
       d\hat{k}^+
       \prod_{j=1}^{N_u-1} \int \frac{d\hat{l}_j^- d^4\hat{l}_{jT}}{(2\pi)^52\hat{l}_j^-}
        \frac{1}{2\hat{l}^-_{N_U}} 
        2\pi\, \delta\!\left( \sum_j \hat{l}_j^+ -q^+ -\hat{k}^+ \right)
        \left(\prod_{j=1}^{N_u} \frac{\hat{l}_j^-}{l_j^-} \right) 
\nonumber \\
  &=&
        d\hat{k}^+ dL(\hat{\underline{l}}_u;q+\hat{k},0) 
        \left(\prod_{j=1}^{N_u} \frac{\hat{l}_j^-}{l_j^-} \right) .
\end{eqnarray}
Hence the Jacobian is
\begin{equation}
\label{eq:jac}
   \Delta_{N_u}(\underline{l}_u,k)
  \equiv  
    \frac{ d\hat{k}^+ dL(\hat{\underline{l}}_u;q+\hat{k},0) }
         { dk^+ dL(\underline{l}_u;q+k;\underline{M}) }
  = \prod_{j=1}^{N_u}\frac{l_j^-}{\hat{l}_j^-}
  = \prod_{j=1}^{N_u}\frac{l_j^-}{l_j^- -k^-/N_u}.
\end{equation}
In the collinear limit for the initial-state, where $k^-$ and $\T{k}$
approach zero, the Jacobian approaches unity.

%---------------------------
\subsection{Construction and proof of factorization}

The proof of factorization proceeds as in \cite{phi3}, with very few
changes.  We first define the contribution $C_R(\Gamma)$ associated with a
leading region $R$ of a graph $\Gamma$, using the operation $T_R$ and some
subtractions.  Then we show that with this definition, the sum over
$R$ of $C_R(\Gamma)$ gives a good approximation to the graph everywhere,
i.e., the remainder in Eq.\ (\ref{eq:regions}) is power suppressed.
Then we sum over graphs for the cross section.  Using the structure of
$T_R\Gamma$ and of $C_R(\Gamma)$ we show that this gives a factorization.  For
the most part it will be sufficient to indicate the differences from
\cite{phi3}.

To allow factorization to be a practical tool for calculations, we
need the hard-scattering factor $\hat{H}$ to be perturbatively
calculable, and we need to construct an algorithm for showering with
perturbatively calculable evolution.  Perturbative calculability of
$\hat{H}$ follows just as \cite{phi3}, since we have the same
subtractive structure; if we set the renormalization and factorization
scales $\mu$ and $\mu_J$ to be of order $Q$, then the integrals for
$\hat{H}$ are dominated by hard momenta on a scale $Q$.  We will
construct a suitable showering algorithm in Sec.\ \ref{sec:alg}.
Before we do this, we will, in Sec.\ \ref{sec:pcf}, analyze the properties of the
parton correlation functions; this
analysis re-uses the methods for factorization of the cross section.

\paragraph{Term for a given region}

As in \cite{phi3}, each leading region is labeled by a particular
pinch-singular surface (PSS) \cite{sterman} in a massless theory, and,
as illustrated in Fig.\ \ref{fig:dis2}, a leading region $R$ of a
graph $\Gamma$ is characterized by a decomposition into subgraphs: a
hard subgraph, one initial-state ``jet subgraph'', and one or more
final state ``jet subgraphs''.  We defined an approximation $T_R$
appropriate to the region in Eq.\ (\ref{eq:approx1}); its action can be
symbolized as converting
\begin{equation}
   \Gamma =
   F_R \times H_R \times 
   U_{R1} \times U_{R2} \times \cdots \times U_{RN_u}
\end{equation}
to 
\begin{equation}
T_R\Gamma = 
   F_R \times (T_R H_R) \times 
   U_{R1} \times U_{R2} \times \cdots \times U_{RN_u} \times \Delta_{N_u} .
\end{equation}
For the representation of the graph as a sum over contributions from
regions, we will define the contribution of region $R$ by applying
$T_R$ to the graph, but only after subtracting the contributions from
smaller regions, to avoid double counting.  As in \cite{phi3}, we
slightly modify this, to include a ``veto factor'' $V_R$:
\begin{equation}
\label{eq:fac-app}
    C_R(\Gamma) = V_R T_R \left( \Gamma - \sum_{R'<R} C_{R'} \Gamma \right).
\end{equation}
These equations give a recursive definition of the contribution
$C_R(\Gamma)$ for region $R$, which starts from the smallest regions.  For
a minimal region, $R_{\rm min}$ i.e., one with no subregions, the
definition is simply $C_{R_{\rm min}} = T_{R_{\rm min}} \Gamma$.

\paragraph{Veto factor}

The veto factor arises because \cite{phi3} the integrals in the cross
section include regions in the hard scattering where partons are
collinear.  Now the subtractions in the hard scattering imply that the
corresponding contributions to the cross section are power suppressed.
Thus in infra-red- and collinear-safe cross sections, the veto factor
is unnecessary.  But the contributions are not completely well
behaved, so we define a veto factor, to eliminate small regions around
the singular surfaces for the massless hard scattering, as follows:
\begin{equation}
        V_R = \prod_S \left[ V(\tilde{M}^2_S/m^2)\,\, 
                       V(-\tilde{k}^2_S/m^2)
                \right].
\end{equation}
The product is over all subsets $S$ of the (massless) final-state
parton lines of the hard scattering.  In this formula $\tilde{M}_S$ is
defined to be the invariant mass of these lines, i.e.,
\begin{equation}
  \tilde{M}^2_S = \left( \sum_{j\in S} \hat{l}_j \right)^2 \geq 0,
\end{equation}
while $\tilde{k}^2_S$ is defined to be the momentum transfer with
respect to the incoming parton line:
\begin{equation}
  \tilde{k}^2_S = \left( \hat{k} - \sum_{j\in S} \hat{l}_j \right)^2
  \leq 0 .
\end{equation}
The elementary veto factor is defined as
\begin{equation}
\label{eq:veto}
        V(M^2/m^2)= \theta(M-m).
\end{equation}
Thus the veto factor $V_R$ for region $R$ eliminates a small
neighborhood of the initial-state and final-state collinear
singularities associated with the massless partons in the hard
scattering.  The exact definition is irrelevant.

We will show an example of the initial-state veto factor in 
Sec.\ \ref{sec:eg}.

\paragraph{Factorization}

Now that appropriate definitions have been made, the proof of
factorization given in \cite{phi3} applies to the case of DIS.  When
we construct $C_R(\Gamma)$, the subtractions are for smaller regions.  As
in \cite{phi3}, a smaller region implies larger collinear subgraphs
and therefore a strictly smaller hard subgraph.  Therefore the
subtractions are applied only inside the hard subgraph:
\begin{equation}
\label{eq:fact.CR}
    C_R(\Gamma) 
  = 
    F_R \times 
    V_R T_R \left( H_R - \sum_{R'<R} C_{R'} H_R \right) 
    \times U_{R1} \times \cdots \times U_{RN_u} \times \Delta_{N_u} .
\end{equation}
This contrasts with the case of gauge theories, where the presence of
leading effects due to soft gluons gives a more complicated
situation, which needs disentangling with the aid of Ward identities. 

The factorization formula is now obtained by summing over all regions
and graphs.  The correspondence between regions and subgraphs converts
the sum to independent sums over graphs for $F_R$, $H_R$, $N_u$ and
the $U_{Rj}$s.  Hence
\begin{equation}
\label{eq:fact1a}
   \sigma[W]
=
   K \sum\int
   \hat{H}_{N_u} \,\,
   \Delta_{N_u} \,\,
   F(\underline{p}_0,P) 
   \prod_{j=1}^{N_u}  \left[ J(\underline{p}_j) \, \Theta(M_j^2/\mu_J^2) \right]
   W(f),
\end{equation}
up to power-suppressed corrections, as usual.  The sum and integral
represent a sum and integral over final states and then a sum over
decompositions of the partonic structure and over the assignment of
final state particles to the $F$ and $J$ factors.  Explicitly, the
factors are as follows:
\begin{enumerate}
\item The hard factor $\hat{H}_{N_u}$ is obtained by summing the hard
  scattering factor in Eq.\ (\ref{eq:fact.CR}) over all graphs for
  $H_R$:
  \begin{equation}
  \label{eq:hard}
       \hat{H}_{N_u}\boldsymbol( \hat{k}, \underline{\hat{l}_u}; 
                   Q, \mu_J, \mu, g(\mu) 
                   \boldsymbol)
     = \sum_{\text{graphs}} V \,T 
       \left(
          H_R - \sum_{R'<R}C_{R'}H_R
       \right),
  \end{equation}
  with a given number of final-state partons.  The relevant graphs are
  for an amputated amplitude times complex conjugate amplitude with
  $N_u$ final-state parton lines, one incoming parton line, and one
  virtual photon.  The unsubscripted $T$ denotes the operation of
  projecting the external partons onto massless momenta and setting to
  zero the mass $m$ on internal lines, while $V$ without a subscript
  is the corresponding veto factor.  The proof in \cite{phi3} shows
  that $\hat{H}$ has no collinear divergences.
  
\item The initial-state factor $F$ is the sum over graphs of $F_R$ for
  a given final-state.  Thus we have an operator formula
  \begin{equation}
    \label{eq:F.def}
      F( \underline{p}_0,P; \mu ) 
    = 
      \left| \langle P | \phi(0) | \underline{p}_0 \rangle \right|^2.
  \end{equation}
  The integral of $F$ over all final states with a given value of $k$
  ($k=P-\sum_{i}p_{0,i}$) gives the pCf
  \begin{equation}
    \label{eq:F.sum}
    \sum_{n_0=1}^{\infty} \int dL(\underline{p}_0;P-k,m_{\rm ph}) \,\, F(\underline{p}_0,P)
    = \Phi(k,P).
  \end{equation}

\item
  Similarly we have a final-state factor for each final-state parton,
  which we define exactly as in \cite{phi3}:
  \begin{equation}
    \label{eq:dU.def}
      J( \underline{p}_j; \mu ) 
    =
      \left| \langle 0 | \phi(0) | \underline{p}_j \rangle \right|^2.
  \end{equation}
  It obeys a sum rule
  \begin{equation}
    \label{eq:dU.sum}
    \sum_{n_j} \int dL(\underline{p}_j;l_j,m_{\rm ph}) 
         \,\, J(\underline{p}_j)
    =
    J_I( l_j^2; \mu ) .
  \end{equation}
\end{enumerate}

We now reorganize the factors in Eq.\ (\ref{eq:fact1a}) to be suitable
for a MCEG.  We assume that we have generated values of the usual $x$
and $Q$ from standard factorization (\ref{eq:fact.incl}), these
variables determining the lepton kinematics.  The problem is now to
find probability densities for the internal variables so that the
generated events give correct weighted exclusive cross sections.  So we first
normalize the factorization formula for $\sigma[W]$ by dividing by the
inclusive cross section $\sigma[1]$.  Then we choose as independent
variables the massless variables, $\hat{k}$ and $\hat{l}_j$, together
with the variables $k^-$, $\T{k}$, and $M_j$ that characterize the
deviation from exact collinearity.  We write a factor that can be used
to generate a hard scattering in terms of the massless projected
momenta independently of the showering. 
Then we write a factor that
gives an approximately normalized\footnote{In a sense to be discussed
below.} 
probability density for $k^-$, $\T{k}$,
and factors that give normalized probability densities in $M_j$, and
for the final states resulting from each final-state parton.  Finally
we insert a factor that 
expresses the deviation of the true formula from an exactly factored
form.  We find
\begin{equation}
\label{eq:fact1b}
\begin{split}
  \frac{ \sigma[W] }{ \sigma[1] }
  ={}&
    \sum_{N_u =1}^{\infty} 
    \int \frac{ d\hat{k}^+ \, dk^- \, d^4\T{k} }{ (2\pi)^6 }
   \int_{ \hat{\underline{l}}_u } \frac{ d\hat{\sigma}_{N_u}\!(Q,\mu_J,\mu) }{ \sigma[1] } \,\,
   f(\hat{k}^+/P^+;\mu) \,\,
   \frac{ \Phi\boldsymbol( ( \hat{k}^+, k^-, \T{k} ) ,P; \mu \boldsymbol) }
        { f( \hat{k}\strut^+/P^+; \mu ) }
   \times
\\
  & \times
   \prod_{j=1}^{N_u} 
      \left[\int_0^\infty \frac{dM_j^2}{2\pi} \,
            \frac{ J_I(M_j^2; \mu) \,\Theta(M_J^2/\mu_J^2) }{ Z(\mu_J/\mu,m/\mu) }
      \right]
   \times
\\
  & \times 
  \left[
      \sum_{n_0} 
      \int \frac{ dL(\underline{p}_0; P-k; m_{\rm ph}) \,\,F(\underline{p}_0,P; \mu) }
             { \Phi(k,P; \mu) }
   \right]
   \prod_{j=1}^{N_u}
       \left[
           \sum_{n_j} \int dD( \underline{p}_j; l_j; \mu ) 
       \right]
   \times
\\
  & \times
   \, \frac{ \Phi(k,P; \mu) }
           { \Phi\boldsymbol( (\hat{k}^+,k^-,\T{k}), P; \mu \boldsymbol) }
   \, W(f).
\end{split}
\end{equation}
Here $f$ is the ordinary renormalized parton density, discussed in
Sec.\ \ref{sec:pdf} below.  
Notice that we have defined $\Phi$ by an operator matrix element in the exact
theory, in which the particle mass is nonzero. Thus the integral over
$k$ is free of the collinear divergences that arise in 
conventional calculations in a massless theory.\footnote{We only make a
   massless approximation in the
   fully subtracted hard-scattering coefficients, where the approximation
   is valid.}  
In addition, $k$ is the exact quark momentum, so that the upper limits
on its components are set by the kinematics of the process, and there
are no ultra-violet problems.  Thus the integral is convergent.

In the first line we use what we will call the differential
hard-scattering cross section defined by
\begin{equation}
\label{eq:sigmah}
    d\hat{\sigma}_{N_u}
  =
     K \, 
     \, dL(\underline{\hat{l}}_u;q+\hat{k},0)\,
     \hat{H}_{N_u}(Q,\mu_J,\mu) \, Z(\mu_J/\mu,m/\mu)^{N_u}.
\end{equation}
As in \cite{phi3}, we have inserted a factor $Z$ for each outgoing
parton.  This provides a kind of analog to the propagator residue
factors in the LSZ reduction formula, and is defined by
\begin{align}
  \label{Z.def}
    Z(\mu_J/\mu,m/\mu) 
  & = 
      \int_0^\infty \frac{dM_j^2}{2\pi} \, \Theta(M_j^2/\mu_J^2) \, J_I(M_j^2;\mu) 
\\
  & =
   \int_0^\infty \frac {dM_j^2}{2\pi }
   \sum_{n_j} \int dL(\underline{p}_j;l_j) \Theta(M_j^2/\mu_J^2) \,
            J(\underline{p}_j;\mu) .
\end{align}
Observe that \cite{phi3} since the integral of $J_I$ over all $M_j$
diverges at large $M_j$, we have to use a cutoff, the factorization
scale $\mu_J$, to define a suitable finite quantity.
We have implemented the cutoff by the function $\Theta(M_j^2/\mu_J^2)$ rather
than by the upper limit of integration, to allow for the possibility
of a smooth cutoff.

But there appears to be no unambiguously appropriate analog of the
factor $Z$ for the incoming parton.  Instead, on the first line of Eq.\ 
(\ref{eq:fact1b}), we have multiplied
$d\hat\sigma$ by the ordinary parton density $f(\hat{k}^+/P^+;\mu)$ taken at
the approximated parton momentum.  This is like the parton density times
hard-scattering in the conventional formalism, and we can use the
numerical values of $d\hat\sigma f$ for generating the kinematic variables
$\hat{k}^+$ and $\hat{l}_j$.  However, as we have already argued, the
value of $\hat{k}^+$ does not agree with the correct value of true
parton momentum.  Since the correct value cannot be known
algorithmically until the full parton kinematics are known, we insert
a correction factor on the last line of Eq.\ (\ref{eq:fact1b}) to
adjust the full cross section to its correct value and we will
implement the correction factor as a reweighting.  

To a first approximation, the incoming parton for the hard scattering
has low transverse momentum and virtuality, and then $k^+$ is close to
$\hat{k}^+$.  So $d\hat\sigma \, f / \sigma[1]$ is approximately a normalized
distribution in $\hat{k}^+$ and the outgoing parton momenta.  Hence the
reweighting corrections are mild.

The remaining factor on the first line, $\Phi/f$, is therefore also
approximately a normalized probability density in $k^-$ and $\T{k}$,
with upper limits determined kinematically.
The density $f$, conventionally called an ``integrated parton
density'', is therefore approximately an integral over $\Phi$.  However,
because of complications from ultra-violet behavior these statements
are not exact;
the detailed relations between $\Phi$ and $f$ will be explored in 
Sec.\ \ref{sec:pcf}.  Again, reweighting corrections to compensate for
slightly inaccurate normalization are mild.

In the second line, the distribution in outgoing parton mass is given
by the $J_I/Z$ factors.  Given our cutoff at $M_j=\mu_J$, this is a
correctly normalized conditional probability density.

The normalized differential distribution of final-states in an
out-going parton given the parton's exact momentum is the quantity
$dD$ defined in Eq.\ (6.42) of \cite{phi3}.  In our current notation
it is
\begin{equation}
  \label{eq:dD.def}
  dD( \underline{p}_j; \mu ) 
  =
  \frac{ J(\underline{p}_j;\mu) \, dL(\underline{p}_j;l_j) }
       { J_I(l_j^2;\mu) }
\end{equation}
The corresponding normalized initial-state factor is
$dL(\underline{p}_0)F/\Phi$.

The deviation from the factorized form of the probabilities is given
by the ratio of the pCf at the correct incoming parton momentum to the
value at the approximated (but off-shell) momentum, i.e., by
the factor $\Phi(k,P) / \Phi\boldsymbol( (\hat{k}^+,k^-,\T{k}),P
\boldsymbol)$.  This factor approaches unity in the collinear limit.
In a MCEG it can be implemented by either weighting generated events by
the factor or using a suitable veto strategy.  Because the
weighting factor is well behaved in the collinear limit, the veto
method is reasonable.  The important point for the logic is that this
reweighting factor is used only within the generation of the hard
scattering, which in our method includes the generation of the
exact off-shell momenta of the external partons of the hard scattering.

\subsection{Choices of $\mu$ and $\mu_J$}
\label{sec:scale.choice}
The renormalization scale $\mu$ and the factorization scale $\mu_J$ are
artifacts of our method of treatment of the theory, so that the cross
section does not depend on them.  As usual, we exploit the dependence
on these variables of factors in factorization theorems to enable
perturbative calculations to be done without large logarithms.  Scale
dependence cancels in predictions of cross sections up to the effect
of uncalculated higher order corrections.

First, the inclusive cross section is calculated from the standard
factorization formula Eq.\ (\ref{eq:fact.incl}) with $\mu$ of order $Q$;
then the coefficient function is a weak-coupling expansion in $g(\mu)$
without large logarithms.  The parton density at scale $\mu \sim Q$ is
related to the parton density at a fixed scale $Q_0$ by use of the
ordinary DGLAP evolution.  The evolution is perturbative.  Our
approach is to define the parton density by \MSbar{} renormalization,
so only the renormalization scale $\mu$ appears here; there is no
separate factorization scale.

Next in the generation of an event is the use of Eq.\ 
(\ref{eq:fact1b}) to give the kinematic distribution of the hard
scattering.  Here we also set $\mu$ and $\mu_J$ of order $Q$ so that the
hard scattering is perturbative.  For consistency we must use the same
values everywhere in Eq.\ (\ref{eq:fact1b}).  The pCf $\Phi$ is an
operator matrix element, so it depends on $\mu$ but not $\mu_J$.  Its
calculation and scale-dependence will be considered in Sec.\ 
\ref{sec:pcf}.

The treatment of the distribution of parton mass $J_I/Z$, on the
second line of Eq.\ (\ref{eq:fact1b}), is exactly as in \cite{phi3}.
Since $\mu$ and $\mu_J$ are the same as in the rest of the formula, they
are of order $Q$.  The calculation of $Z(\mu_J/\mu,m/\mu)$ is then
perturbative, in powers of $g(\mu)$, with the mass dependence being
negligible.  But $J_I(M_j^2; \mu)$ will have logarithms of $M_j/\mu$,
which get large\footnote{Note that as a simple operator matrix element
  $J_I$ has no $\mu_J$ dependence.} when $M_j\ll Q$.  Its evolution is
governed by the anomalous dimension of the parton field and therefore
by the same RGE as the pCf, Eq.\ (\ref{eq:pcf.RG}) below.  Thus we use
the renormalization group to express it in terms of $J_I$ at another
scale:
\begin{equation}
  J_I(M_j^2; \mu)
  =
  \exp\left[ - \int_{\mu'}^\mu \frac{dv}{v} \gamma\boldsymbol( g(v) \boldsymbol) \right]
  J_I(M_j^2; \mu').
\end{equation}
When $M_j$ is in a perturbative region, we set $\mu'$ of order $M_j$, so
that $J_I(M_j^2; \mu')$ can be computed perturbatively in powers of
$g(\mu')$. When $M_j$ is too small, we set $\mu'$ to a fixed scale $Q_0$
and use a non-perturbative model (fit to data).  The model is
constrained by a sum rule that $J_I$ integrates to the perturbatively
calculable quantity $Z$.  

In the third line of Eq.\ (\ref{eq:fact1b}), we have factors that give
the distribution of hadronic states given the exact parton
momenta.  Each factor is a phase-space differential times a ratio of
two matrix elements.  The matrix elements differ only by whether or
not they are integrated over final states, so the numerator and
denominator have the same anomalous dimension.  Thus we can replace
$\mu$ separately in each factor by a scale appropriate to each parton:
of order $M_j$ for a final-state parton or of order $\sqrt{|k^2|}$ for
the initial-state parton.  (A fixed scale would be more appropriate in
the non-perturbative region of momenta.)  Then we exploit
factorization of the pCf to further the calculation.  The change in
the value of the scale $\mu$, potentially very different in different
factors, is the key feature that enables the MCEG approach to give
systematically reliable estimates of cross sections.

The final factor 
$\Phi(k,P) / \Phi\boldsymbol( (\hat{k}^+,k^-,\T{k}), P\boldsymbol)$
 is again a
ratio of quantities with the same anomalous dimension, so that it is
scale-independent.  The values needed are obtained from the
calculation of the pCf.

\subsection{Collinear safety, renormalization group equations}

For the factorization formula to have predictive power we need the
following:
\begin{itemize}
\item The remainder, i.e., the difference between the factorized
  formula for the cross section and the true cross section, is power
  suppressed.
\item Infrared and collinear safety of the subtracted hard scattering
  factors $\hat{H}$.
\item Infrared safety of the integrated final-state jet factors.
\item Renormalization-group-like equations for the scale dependence of
  the jet factors.
\item Theorems for the parton correlation functions.
\end{itemize}
The power suppression of the error and the collinear and infra-red
safety follow exactly as in \cite{phi3}, and the evolution equations
for the final-state jet factors are the same.  Then as explained in
the previous section, Sec.\ \ref{sec:scale.choice}, perturbatively
based calculations can be made by combining suitable choices of $\mu$
and $\mu_J$ in each factor with the evolution equations to relate the
factors to their values with a common choice of scale.

\subsection{Factorization schemes}
\label{sec:schemes}

We have a choice in the algorithm for generating parton configurations
for the hard scattering:
\begin{itemize}
\item[(a)] We can follow the structure of Eq.\ (\ref{eq:fact1b}): (i)
  Generate a conventional massless hard scattering. (ii) Generate
  values for $k^-$, $\T{k}$, and the $M_j$'s. (iii) Compute the exact
  parton momenta $k$ and $\underline{l}_u$. (iv) Reweight by the ratio
  of the correct pCf to the approximated pCf.

Or:
\item[(b)] We can generate the hard scattering according to a formula
  in terms of the parton correlation functions alone with exact parton 
  momenta.
\end{itemize}
In both cases initial-state and final-state showering are performed
conditional on the exact parton momenta, so the change in algorithm
affects only the algorithm for the hard scattering, not the rest of
the MCEG or the probabilities.  For the second case, we can exhibit
the necessary probability densities by rewriting Eq.\ 
(\ref{eq:fact1b}) in a form closer to that of Eq.\ (\ref{eq:fact1a}):
\begin{equation}
\label{eq:fact1c}
\begin{split}
  \frac{ \sigma[W] }{ \sigma[1] }
  ={}&
    \sum_{N_u =1}^{\infty} 
       \prod_{j=1}^{N_u} 
            \left[ \int \frac{ d^6l_j }{ (2\pi)^6 } \, J_I(l_j^2) \, \Theta(l_j^2/\mu_J^2)
            \right]
  \frac{ K \, \hat{H}_{N_u} \, \Phi(k,P)\, \Delta_{N_u}(\underline{l}_u, k)  }
       { \sigma[1] } \times
\\
  & \times 
  \left[
      \sum_{n_0} 
      \int \frac{ dL(\underline{p}_0; P-k; m_{\rm ph}) \,\,F(\underline{p}_0,P) }
             { \Phi(k,P) }
   \right]
   \,
   \prod_{j=1}^{N_u}
       \left[
           \sum_{n_j} \int dD( \underline{p}_j; l_j; \mu ) 
       \right]
\\
  & \times
   \, W(f),
\end{split}
\end{equation}
with $k=\sum_jl_j-q$.  

Which of these algorithms will be better in practice is not yet
clear.  That will depend on practical experience, at least.

But clearly, if we use the second method, it would appear more natural
to remove the Jacobian factor $\Delta_{N_u}$ from Eq.\ (\ref{eq:fact1c}).
If we did this, then consistency between the two event-generation
methods requires us also to \emph{insert} a factor of $1/\Delta_{N_u}$ in
the formula (\ref{eq:fact1b}) used to specify the probabilities in the
first scheme.  We would also need to remove the factor $\Delta_{N_u}$ from
Eq.\ (\ref{eq:fact1a}).  Consistency with the derivation of these
equations requires a corresponding change in the definition of $T_R$:
we would insert a factor $1/\Delta_{N_u}$ in Eq.\ (\ref{eq:approx}) and
remove a factor $\Delta_{N_u}$ from Eq.\ (\ref{eq:approx1}).

Since $T_R$ changes, we would find that changes in the higher-order
terms in the hard-scattering coefficients occur, as is forced by the
subtractions in the recursive definition (\ref{eq:fac-app}) of the
contribution $C_R$ of an individual region.  This gives us two
schemes, each more natural to one of the two event generation
algorithms.  As usual, changes in the scheme are compensated in
higher-order corrections to the hard scattering, so that the physical
cross sections are invariant up to uncalculated corrections of yet
higher order.

%==========================================================
\section{Parton correlation functions (pCf's)} 
\label{sec:pcf}

To build an EG for DIS we will need the values of the pCf $\Phi(k,P)$, to
use in the hard-scattering factorization formula.  At large $k^2$,
this will involve a new factorization formula that expresses it in
terms of ordinary parton densities (defined, say, in the \MSbar{}
scheme).  This is exactly analogous to the standard factorization
property for the cross section, Eq.\ (\ref{eq:fact.incl}).  The
showering of the parton, and hence the analysis of the final states
inside $\Phi$, will need a second factorization formula analogous to Eq.\ 
(\ref{eq:fact1b}) or (\ref{eq:fact1c}) for the cross section.

The second factorization formula for the pCf concerns a quantity
$\Phi[W]$ whose 
relation to the pCf $\Phi$ is like that of the weighted cross section
$\sigma[W]$ to the unweighted DIS cross section $\sigma[1]$.  To be precise,
we define
\begin{equation}
\label{eq:pcf.W}
   \Phi_{[W]}(k,P;\mu)
   = \sum_{n_0}
    \int dL(\underline{p}_0; P-k; m_{\rm ph}) \,\,
    F(\underline{p}_0,P;\mu) \,
    W(f),
\end{equation}
where $F$ is the same as in Eq.\ (\ref{eq:F.def}).  
Of course, the previously defined unweighted pCf, Eq.\,(\ref{eq:F.sum}),
is the weighted pCf
with unit weight for all states: $\Phi(k,P)=\Phi_{[1]}(k,P)$.

This factorization formula for the weighted pCf implies an integral
equation for the unweighted pCf in terms of itself, so that it is not
an easy equation to solve for $\Phi$.  However the first
factorization property for the pCf gives the unweighted $\Phi$ in terms
of ordinary parton densities, 
and it therefore gives an effective method for calculating its functional
form.  Given this information, the second factorization is perfectly
adapted to analyzing the partonic content of the final states, i.e.,
as a basis for the showering of an initial-state parton.  In addition
we need renormalization group equations so that the renormalization
scale $\mu^2$ can be chosen of order $|k^2|$ to exploit the
factorization theorems perturbatively at large $|k^2|$.

The uses of the factorization theorems at large $|k^2|$ are
perturbative, aided by renormalization-group transformations.  But at
low $|k^2|$ in an asymptotically free theory, we must resort to
non-perturbative methods, modeling and simple data fitting, just as
in current MCEGs.

\subsection{Renormalization group}
\label{sec:pcfre}
Let us first observe that the pCf (weighted or unweighted) arises from
matrix elements of two renormalized fields, so it has a
renormalization-group equation
\begin{equation}
\label{eq:pcf.RG}
   \frac{d}{d\ln\mu} \Phi( k,p; \mu )    
   = -\gamma\boldsymbol( g(\mu) \boldsymbol) \, \Phi,
\end{equation}
with the conventions of \cite{jccbook}.  As usual, the total
derivative applies to both the direct $\mu$-dependence of the Green
function and the $\mu$-dependence of the running coupling and mass.
This formula enables us to transform the pCf from $\mu\simeq Q$ as used in
hard scattering factorization to the value $\mu^2\simeq|k^2|$ as used when
calculating the value of $\Phi$ from the first factorization or when
applying the second factorization for showering.

\subsection{Standard parton densities}
\label{sec:pdf}
The definition of the standard \MSbar-pdf in $\phi^3$ theory is given in
\cite{factn,f2}.  Here we review it and some of the properties of the pdf.
First there is the unrenormalized pdf:
\begin{equation}
        f_0(x) = xP^+ \int dy^- \, e^{-ixP^+y^-} \,
                \langle P| \, \phi_0(0,y^-, \T{0}) \, \phi_0(0) \, |P\rangle_c.
\end{equation}
The subscript ``$c$'' on the matrix element means that we take only
the component in which the fields are connected to the target
particle: the vacuum expectation value is subtracted out.  Implicitly,
$f_0$ has dependence on the parameters of the theory, $g(\mu)$,
$m(\mu)$ and $\mu$, as well as on the UV regulator parameter, which we
take to be the dimension $n$ of space-time.  Notice that we use bare
fields in this formula, which implies that there is an ultra-violet
divergence caused by that in $Z_2$.  But, if in an
attempt to obtain a UV-finite renormalized parton density, we
simply replaced the bare fields by renormalized fields, that would not
work because there is a
further UV divergence caused by the unrestricted integral over the
transverse momentum and virtuality of the external parton.  So the
\MSbar{} pdf is defined to have all the UV divergences renormalized.
It has the form 
\footnote{The necessary dependence of the bare parton density on the
  the regulator parameter $n$, both directly and through the
  $n$-dependence of the bare coupling and mass, has not been indicated
  explicitly in this formula.} 
\begin{eqnarray}
     f(x;\mu) &=& \lim_{n\to6} \int_1^{x^{-1}} \frac{d\eta}{\eta} K(\eta,g,n) f_0(\eta x)
        \nonumber \\
        K(\eta,g,n) &=& \delta(\eta-1) + \sum_{N=1}^{\infty} (n-6)^{-N}
           K_N( \eta, g ),
\end{eqnarray}
where $K(\eta,g,n)$ is a series of pure poles at the physical space-time
dimension $n=6$.

Notice that in the absence of these divergences, the bare parton
density would be an integral over the pCf with a factor $Z_2$:
\begin{align}
  f_0(x) = Z_2 \int \frac{ dk^- \, d^4\T{k} }{ (2\pi)^5 } \, \Phi(k,P).
\end{align}
Since we can analyze issues of UV behavior perturbatively, this
equation suggests that a suitably improved analysis would relate $f$
to an integral over $\Phi$.  Hence, the factor $\Phi$/f in
Eq.\ (\ref{eq:fact1b}) is approximately a probability distribution in
$k^-$ and $\T{k}$ conditional on the value of $\hat{k}^+$.  Because of
the existence of higher-order corrections to the UV coefficient
functions in relations between $\Phi$ and $f$, we must expect such
statements not to be exact, however.  We will give a detailed analysis
in Sec.\ \ref{sec:pdf.relation}, and merely observe here that, in a
leading logarithm approximation,
we can approximate renormalization by a cut-off.  Thus we can
approximately write the usual pdf at a large scale $M$ as:
\begin{equation}
\label{eq:f.LLA}
f(x;g(M),m,\mu=M) =
\int_{|k^2|, k_T^2<M^2} \frac{ dk^- \, d^4\T{k} }{ (2\pi)^5 } \, \Phi(k,P;M)
+O(g(M)^2).
\end{equation}
It is only in this approximate sense that the usual designation of $f$
as an ``integrated pdf'' is justified.

Observe that the scale dependence of the parton density arises from
two sources: the explicit cutoff on the $k$ integral, and the
anomalous dimension of the renormalized parton field in the operator
definition of $\Phi$.

%------------------------------------------------------------
\subsection{Kinematics of pCf}

In non-gauge theories, by Lorentz invariance, the parton correlation 
function $\Phi(k,P;\mu)$
depends on $k$ only through the scalar variables $k^2$ and $k\cdot P$.
Kinematically $\Phi$ has the same kind of variables as a DIS structure
function, so it is convenient to 
use variables analogous to the standard $x$ and $Q^2$ of DIS:
\begin{equation}
   \zeta \equiv \frac{k^2}{2P\cdot k}, \,\,\,\,\, R^2 \equiv -k^2.
\end{equation}
The kinematics in the factorization formula (\ref{eq:fact1b}) for DIS
correspond to the first graph in Fig.\ \ref{fig:dis2}, so $k$ is
space-like.  Hence, even though the pCf is defined for
all values of $R^2$, it is only used when $R^2$ is positive.  Just as
in DIS, the positivity of the squared mass of the final state implies
that $0<\zeta\leq1$, when $R\gg m$.  Since the kinematics of the pCf are
completely analogous to those of DIS structure functions, so are the
factorization properties, essentially obtained by the replacement of
currents by parton fields. 

In gauge theories like QCD, the gauge-invariance of the theory
requires us to insert Wilson lines in the operator definition of
pCf's.  Appropriate direction(s) for the Wilson lines are those
compatible with a proof of factorization, a proof that would be much
more complicated than in our $\phi^3$ theory.  If $n^\mu$ is the direction
of the Wilson line, then there are two extra scalar variables: $(P\cdot
n)^2/n^2$ and $k\cdot n/P\cdot n$.  (Since $\Phi$ is invariant under scaling of
$n$ only variables invariant under this scaling can be used.)  The
extra two variables will increase the computational 
cost: for example in a much larger interpolation table to be used in
the numerical evaluation of the pCf.  In addition there will be more
complications in the factorization theorems for the pCf.  Note that it
should be possible to obtain the dependence on one of the extra
variables, probably $(P\cdot n)^2/n^2$, by a suitable generalization of
the Collins-Soper equation \cite{TMD}.

\subsection{Relations between ordinary parton density and parton
  correlation function}
\label{sec:pdf.relation}
There are two kinds of relation between the ordinary pdf and the pCf $\Phi$.
The first gives a result for $\Phi$ when the parton virtuality is large.
The second result gives an integral of $\Phi$ over parton momenta $k$.

\subsubsection{PCf at large $|k^2|$}

The result for large $|k^2|$ arises because,
as we have already stated, the pCf is like a structure function but
with different operators.
When $|k^2|$ is much larger then some typical nonperturbative scale,  
it follows that there is a factorization theorem for the unweighted
pCf of the same form as Eqs.\ (\ref{eq:fact.incl}) for the DIS cross
section or structure function:
\begin{equation}
  \label{eq:pcf.fact.incl}
  \Phi(k,P;\mu) = \int_\zeta^1 \frac{d\xi}{\xi} \, 
           C_\Phi\boldsymbol(\zeta/\xi,-k^2/\mu^2,g(\mu)\boldsymbol) \,\, f(\xi;\mu) 
  + \mbox{power-suppressed correction}
  .
\end{equation}

Calculations of the coefficient $C_\Phi$ are completely analogous to
calculations 
of the structure function $F_2$, except for details due to
differences in the graphs.  For example the lowest-order non-trivial
graph for a pCf of a
parton target Fig.\ \ref{fig:treeg} is
\begin{equation}
     \Phi(k,P;\mu)_{LO} 
   =
     2\pi \, \delta\boldsymbol((P-k)^2-m^2\boldsymbol)
        \frac{g(\mu)^2}{ (k^2-m^2)^2 }
   =
     2\pi \frac{\zeta}{|k^2|} \delta(\zeta-1)  \frac{g(\mu)^2}{ (k^2-m^2)^2 }.
\end{equation}
Notice that unlike the case of $f$, which is integrated over
$k^2$, the graph without any interactions does not contribute here
since we will make $|k^2|$ large.
The lowest order term in the coefficient $C_\Phi$ is therefore
\begin{equation}
  \label{eq:C.Phi.LO}
  C_\Phi(\zeta/\xi,R^2;\mu,g) = \frac{ 2\pi g^2 }{ R^6 } \zeta \delta(\zeta-\xi) + O(g^4).
\end{equation}
Hence the lowest order estimate of the unweighted pCf is
\begin{equation}
  \label{eq:pcf.LO}
  \Phi(k,P; \mu) 
  = 
     \frac{ 2\pi g(\mu)^2 }{ |k^2|^3 } 
     f(\zeta;\mu) \,
     + O\left( \frac{ g^4(\mu) }{ |k^2|^3 } \times \mbox{logarithms} \right),  
\end{equation}
\FIGURE{
\centering
\psfrag{k}{\small $k^\mu$}
\psfrag{P}{\small $P^\mu$}
\psfrag{P-k}{\small $P^\mu-k^\mu$}
\includegraphics[scale=.6]{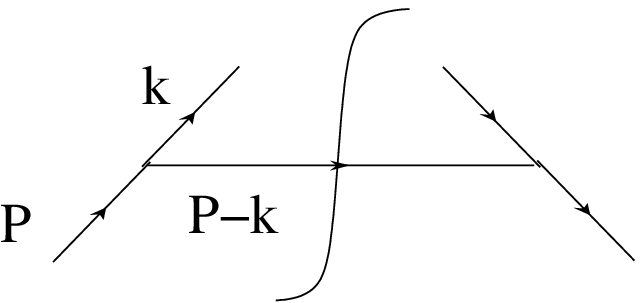}
\caption{Tree graph for $\Phi(k,P)$ of a parton target.}
\label{fig:treeg}
}
\noindent
where, as already mentioned, $\zeta = k^2/(2k\cdot P)$.  
Note, that this formula is only to be applied at large $|k^2|$.  At
low $k^2$ we cannot neglect the parton mass, and therefore the apparent
singularity at $k^2=0$ in Eq.\ (\ref{eq:pcf.LO}) is not physical.

In the higher terms in Eq.\ (\ref{eq:pcf.LO}), there are, as usual in
such expansions, logarithms of $k^2/\mu^2$.  Since we will use $\Phi$ in,
for example, a calculation of hard scattering at scale $Q$, but need
its value when $|k^2|$ is much less than $Q^2$, we must 
evaluate the right-hand side of Eq.\ (\ref{eq:pcf.LO}) after a
renormalization-group transformation using Eq.\ (\ref{eq:pcf.RG}).
Thus practical calculations will use
\begin{equation}
\label{eq:pcf.LO.RG}
\begin{split}
  \Phi(k,P; \mu) 
  = {}&
     \frac{ 2\pi g^2 \boldsymbol( \sqrt{|k^2|}  \boldsymbol) }{ |k^2|^3 } 
     f(\zeta; \sqrt{|k^2|}) \,\,
     \exp\left(
           -
           \int_{\sqrt{|k^2|}}^{\mu} \frac{d\mu'}{\mu'}  
               \gamma\boldsymbol( g^2(\mu') \boldsymbol)
         \right)
\\
   &
     + O\left( \frac{ g^4(|k|^2) }{ |k^2|^3 } \right) .
\end{split}
\end{equation}
On the right-hand side of this formula we have transformed the
renormalization scale to $\sqrt{\strut|k^2|}$.  But the scale may be
multiplied by a factor of order unity, which could be used to optimize
for the likely size of uncalculated higher order corrections.

\subsubsection{Integral of pCf}

The second result for the pCf was proved (with a slight variation) by
one of us in 
\cite{f2}. It says that if we define a quantity $\tilde{f}$ as the
integral of $\Phi$ over $k^-$ and $\T{k}$, with an upper limit on $\T{k}$
and virtuality:
\begin{equation}
\label{eq:f.tilde.def}
 \tilde{f}(x,M;g(\mu),m,\mu) = \int_{|k^2|, k_T^2<M^2} 
 \frac{ dk^- \, d^4\T{k} }{ (2\pi)^5 } \, \Phi(k,P;g(\mu),m,\mu),
\end{equation} 
then this has an expansion in terms of the ordinary parton density:
\begin{equation}
\label{eq:f.tilde.fact}
  \tilde{f}(x,M;g(\mu),m,\mu) = \int_{x}^{1} d\xi \xi^{-1} T(\xi/x, M^2/\mu^2, g(\mu)) f(\xi)
  +\mbox{power-suppressed correction}.
\end{equation}
The formulation in \cite{f2} differs only in that the limit was
imposed only on $\T{k}$ and not on $|k^2|$ as well, but the proof is
unaltered.   The lowest order value of the coefficient $T$ is
$\delta(\xi/x-1)$, so that $\tilde f = f + O(g^2)$.  
From this follows the previously announced result Eq.\
(\ref{eq:f.LLA}); the \MSbar{} pdf is approximately the integral of an
unintegrated parton density.

This result gives a sum rule involving an integral over the whole
non-perturbative region for the pCf, thereby providing a constraint on
modeling of this region given that the ordinary pdf has already
been measured.

\subsection{Recursive factorization for weighted pCf in terms of pCf}
In Eq.\ (\ref{eq:fact1b}) the weighted DIS cross section was expressed
with an integral over exact parton momenta in terms of the unweighted
pCf. With the same reasoning, there is an analogous factorization
property for the (weighted) pCf.  Since we will wish to use it to give
conditional probability densities for the internal partons in a MCEG,
we normalize the formula to the unweighted pCf:
\begin{equation}
\label{eq:pcf.fact.W}
\begin{split}
  \frac{ \Phi_{[W]}(k,P;\mu) }{ \Phi(k,P;\mu) }
  ={}&
    \sum_{N_u =1}^{\infty} 
    \int \frac{ d\hat{r}^+ \, dr^- \, d^4\T{r} }{ (2\pi)^6 }
   \int_{ \hat{\underline{l}}_u } 
      \frac{ d\hat{H}_{\Phi,N_u}(\mu,{\mu'}_J)  }{ \Phi(k,P;\mu) } \,\,
   f(\hat{r}^+/P^+;\mu) \,\,
   \times
\\
  & \times
   \frac{ \Phi\boldsymbol( ( \hat{r}^+, r^-, \T{r} ) ,P ;\mu \boldsymbol) }
        { f(\hat{r}\strut^+/P^+;\mu) }
   \times
\\
  & \times
   \prod_{j=1}^{N_u} 
      \left[\int_0^\infty  \frac{dM_j^2}{2\pi} \,
            \frac{ J_I(M_j^2;{\mu'}_J,\mu)\,\Theta(M_j^2/{\mu'}_J^2) }{ Z({\mu'}_J/\mu,m/\mu) }
      \right]
   \times
\\
  & \times 
  \left[
      \sum_{n_0} 
      \int \frac{ dL(\underline{p}_0; P-r; m_{\rm ph}) \,\,F(\underline{p}_0,P;\mu) }
             { \Phi(r,P;\mu) }
   \right]
   \prod_{j=1}^{N_u}
       \left[
           \sum_{n_j} \int dD( \underline{p}_j; l_j; \mu ) 
       \right]
   \times
\\
  & \times
   \, \frac{ \Phi(r,P;\mu) }
           { \Phi\boldsymbol( (\hat{r}^+,r^-,\T{r}), P ;\mu  \boldsymbol) }
   \, W(f).
\end{split}
\end{equation}
The main change compared with Eq.\ (\ref{eq:fact1b}) is that the
differential hard scattering factor has been given a different name,
$d\hat{H}_\Phi$ instead of $d\hat{\sigma}$.  In addition, the symbol for the
internal loop momentum has been changed from $k$ to $r$, to avoid a
conflict with the symbol for the external parton momentum.  The
differential hard scattering coefficient $d\hat{H}_\Phi$ is defined in
terms of subtracted hard scattering graphs in analogy to $d\hat\sigma$ in
Eq.\ (\ref{eq:sigmah}):
\begin{equation}
\label{eq:dH.Phi}
    d\hat{H}_{\Phi,N_u}(\mu,{\mu'}_J)
  =
     dL(\underline{\hat{l}}_u;k+\hat{r},0)\,
     \hat{H}_{\Phi,N_u}(\mu,{\mu'}_J) \, Z({\mu'}_J/\mu,m/\mu)^{N_u}.
\end{equation}
The approximated internal parton momenta $\hat{r}$ and $\hat{l}_j$ are
defined as before, except for replacing $q$ and $k$ in the definitions
in Sec.\ \ref{sec:proj} by $-k$ and $r$.

Notice how we have used a different symbol ${\mu'}_J$ for the
factorization scale than we did in factorization for the cross
section.  When we employ Eq.\ (\ref{eq:pcf.fact.W}), we will use the
fact that the anomalous dimensions of the numerator and denominator on
the left-hand side are the same, so that we can replace the value $\mu\sim
Q$ used in factorization for the cross section by a value
$\mu\sim\sqrt{\strut|k^2|}$ appropriate for the use of (\ref{eq:pcf.fact.W}).  We
must also choose ${\mu'}_J$ of the same order, so that the
hard-scattering factor in this equation is perturbatively calculable,
without large logarithms of $k^2/{\mu'}_J^2$ or of $k^2/\mu^2$. 

It is also possible to write (\ref{eq:pcf.fact.W}) in terms of unapproximated
momenta, as we did for the cross section in Eq.\ (\ref{eq:fact1c}):
\begin{equation}
\label{eq:pcf.fact.Wa}
\begin{split}
  \frac{ \Phi_{[W]}(k,P;\mu) }{ \Phi(k,P;\mu) }
  ={}&
    \sum_{N_u =1}^{\infty} 
       \prod_{j=1}^{N_u} 
            \left[ \int \frac{ d^6l_j }{ (2\pi)^6 } \, J_I(l_j^2) 
            \right]
    \frac{ \hat{H}_{\Phi,N_u}(\mu,{\mu'}_J) \, \Phi(r,P;\mu)\, \Delta_{N_u}(\underline{l}_u,r) }
         { \Phi(k,P;\mu) } \times
\\
  & \times 
  \left[
      \sum_{n_0} 
      \int \frac{ dL(\underline{p}_0; P-r; m_{\rm ph}) \,\,F( \underline{p}_0,P; \mu ) }
             { \Phi(r,P;\mu) }
   \right]
   \,
   \prod_{j=1}^{N_u}
       \left[
           \sum_{n_j} \int dD( \underline{p}_j; l_j; \mu ) 
       \right]
\\
  & \times
    \, W(f),
\end{split}
\end{equation}
with $r=\sum_jl_j+k$.  

The lowest-order value of $\hat{H}_{\Phi,1}$ is the essentially the same as
the coefficient for the pCf in terms of integrated parton densities:
\begin{equation}
  \label{eq:H.Phi.LO}
  \hat{H}_{\Phi,1}(k,\hat{r}) = \frac{ 2\pi g^2 }{ |k^2|^3 } \delta(-1+2k\cdot\hat{r}/k^2) + O(g^4).
\end{equation}
The lowest-order value of the $\hat{H}_{\Phi,N_u}$ with more external outgoing
partons is of order $g^{2N_u}$.

%----------------------------------------------------------
\section{Implementation}
\label{sec:imp}

%\subsection{The algorithm}
\label{sec:alg}
We now put the information together to give an algorithm for a MCEG.
First, the ordinary factorization formula (\ref{eq:fact.incl}) gives us
a calculation of the inclusive cross section (and hence of the
structure function) as a function of $x$ and $Q$.  Then
(\ref{eq:fact1b}) expresses the weighted cross section in terms of
the pCf.  Exactly similar factorization formulas apply to the pCf:
Eqs.\ (\ref{eq:pcf.fact.incl}) and (\ref{eq:pcf.fact.W}).  Lowest
order values for the coefficients are presented below the
corresponding factorization formulas.

We now assume that we have started to generate an event by generating
values of $x$ and $Q$ according to probabilities given by the 
usual factorization formula for the inclusive cross section.  We wish
to generate a full event.  
\begin{enumerate}
\item Given the momenta of the photon and the hadron, $q^\mu$ and
  $P^\mu$, we generate a value $N_u$ for the number of final-state
  partons and then a final-state partonic configuration $(l_1, \ldots,
  l_{N_u})$ with the density extracted from Eq.\ (\ref{eq:fact1c})
  \begin{equation}
  \label{eq:H.den}
       \prod_{j=1}^{N_u} 
            \left[ \frac{ d^6l_j }{ (2\pi)^6 } \, J_I(l_j^2) 
            \right]
       \frac{ \hat{H}_{N_u} \, \Phi(k,P) }
            { \sigma[1] }
       \Delta_{N_u}.
   \end{equation}
   This can either be done directly, or by separate steps of
   generating a massless configuration $(\hat{l}_1, \ldots, \hat{l}_{N_u})$
   together with values of $k^-$, $\T{k}$ and $M_1$, \ldots $M_{N_u}$, as
   at the end of Sec.\ \ref{sec:rationale}, with probability densities
   extracted from from Eq.\ (\ref{eq:fact1b}).  In the second case we
   need to calculate the true partonic variables.  

   The hard scattering is evaluated perturbatively with the
   renormalization and factorization scales $\mu$ and $\mu_J$ chosen to be
   of order $Q$, to avoid large logarithms.
   
 \item For each final-state parton $j$ that is above some threshold of
   perturbative invariant mass $l_j^2>Q_0^2$, generate a final-state
   parton shower using the algorithm given in \cite{phi3} in its own
   CM frame. Then boost it to the true $l_j$.  Since we already know
   the exact momentum of the parton, we should omit the step of
   finding the invariant mass.

   Otherwise the parton is left unchanged for nonperturbative
   hadronization. 
   
 \item If the initial-state parton is above the threshold of
   perturbative invariant mass $|k^2|>Q_0^2$, generate its parton
   shower. This is done by looping back to the first step, with
   the photon momentum $q$ replaced by the parton momentum $-k$, and
   with the factorization formulas changed to the ones for the pCf.

   Otherwise the parton is left unchanged for nonperturbative
   hadronization. 
   
\item Once all remaining partons that are above the perturbative
   threshold have been showered, apply nonperturbative hadronization.

\end{enumerate}

This recursive algorithm for event generation reduces the whole
problem to a repeated application of the same procedure, so that the
program size is independent of the number of particles generated and
the computational resources are linear in the number of particles.

%==========================================================
\section{Zero- and one-loop corrections to coefficient functions}
\label{sec:eg}

In this section we give the coefficient functions for the structure
function and the parton correlation function up to one-loop order.

%----------------------------------------
\subsection{Structure function}

For the hard scattering coefficients for the DIS structure function we
could work directly with diagrams for the hard-scattering component
of factorization, i.e., diagrams for the subgraph labeled ``H'' in a
decomposition like Fig.\ \ref{fig:dis2}.  However, we will find it
convenient to put the graphs in the context of their use in the
factorization theorem.  Thus we consider graphs of the form of Fig.\ 
\ref{fig:oneloop}, where we have attached a full amplitude for the pCf
$\Phi$ to the graphs from which the one-loop hard-scattering coefficient
function is calculated.  

\FIGURE{
\centering
\psfrag{P}{\small $P$}
\psfrag{k}{\small $k$}
\psfrag{s}{\small $s$}
\psfrag{l1}{\small $l_1$}
\psfrag{l2}{\small $l_2$}
\psfrag{p2}{\small $p_2$}
\psfrag{p1}{\small $p_1$}
\psfrag{k1}{\small $k_1$}
\psfrag{f}{$\Phi(k,P)$}
\psfrag{q}{$q$}
        \begin{tabular}{c@{~~~}c@{~~~}c}
          \includegraphics[scale=.26]{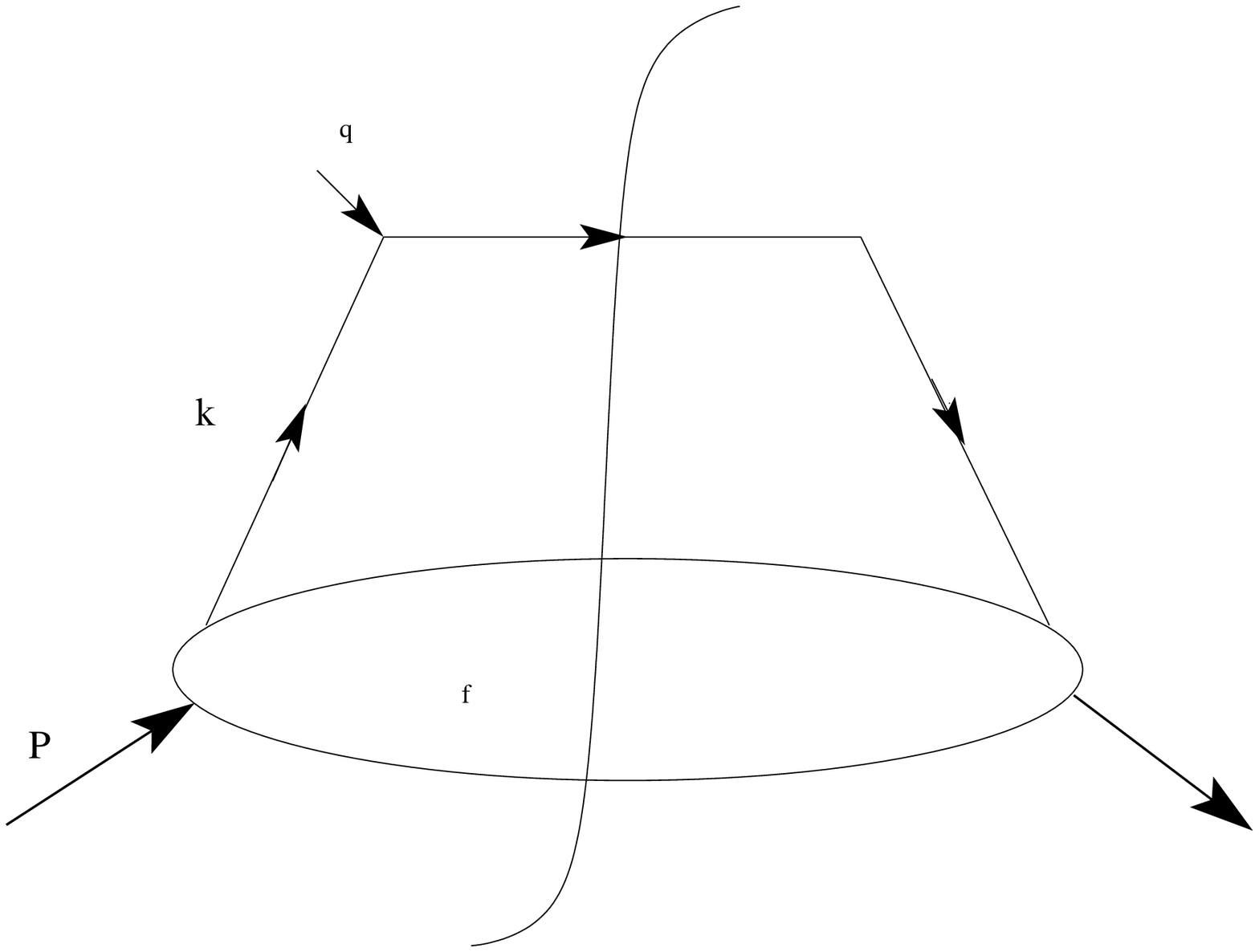}
           &
          \includegraphics[scale=.26]{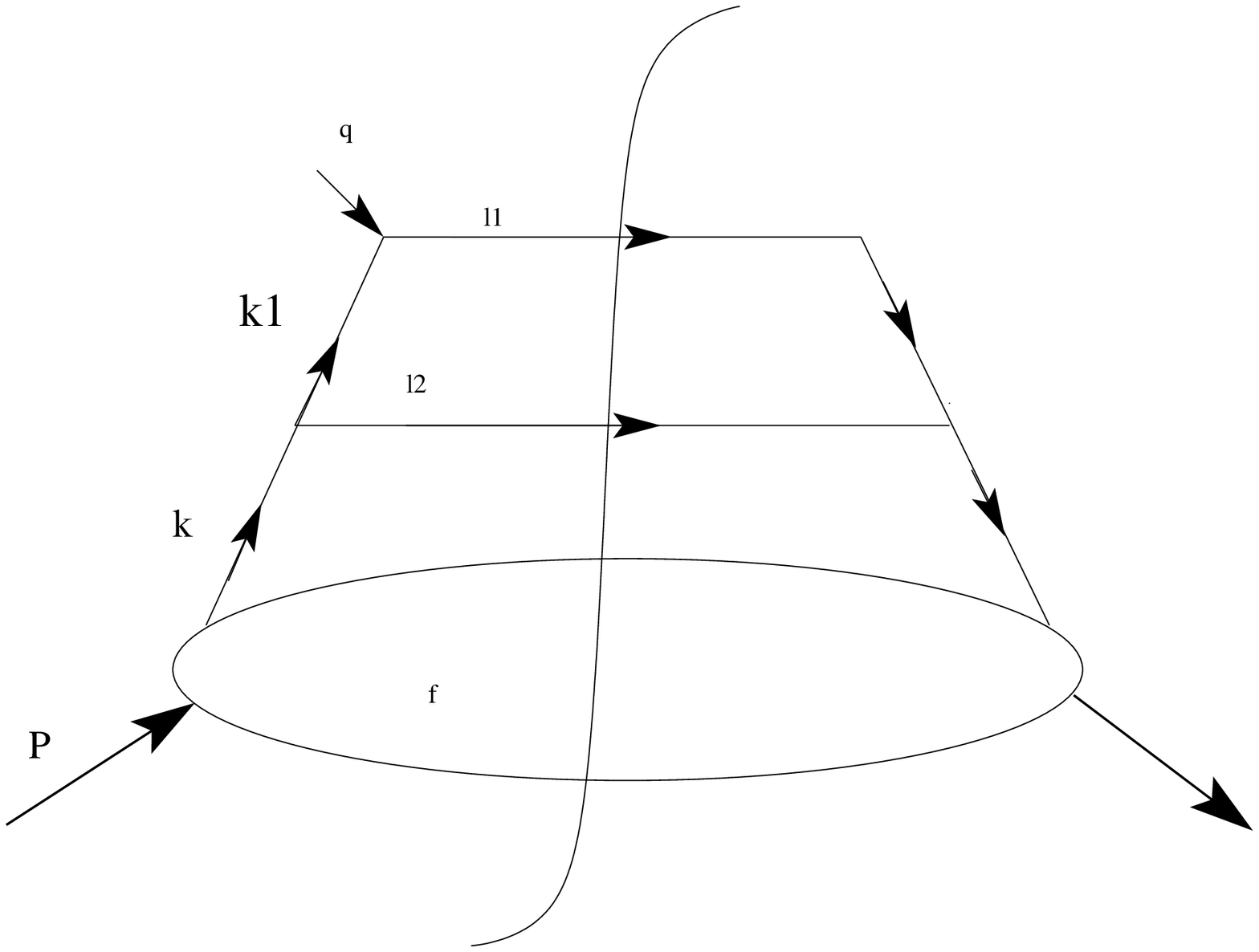}
           &
          \includegraphics[scale=.26]{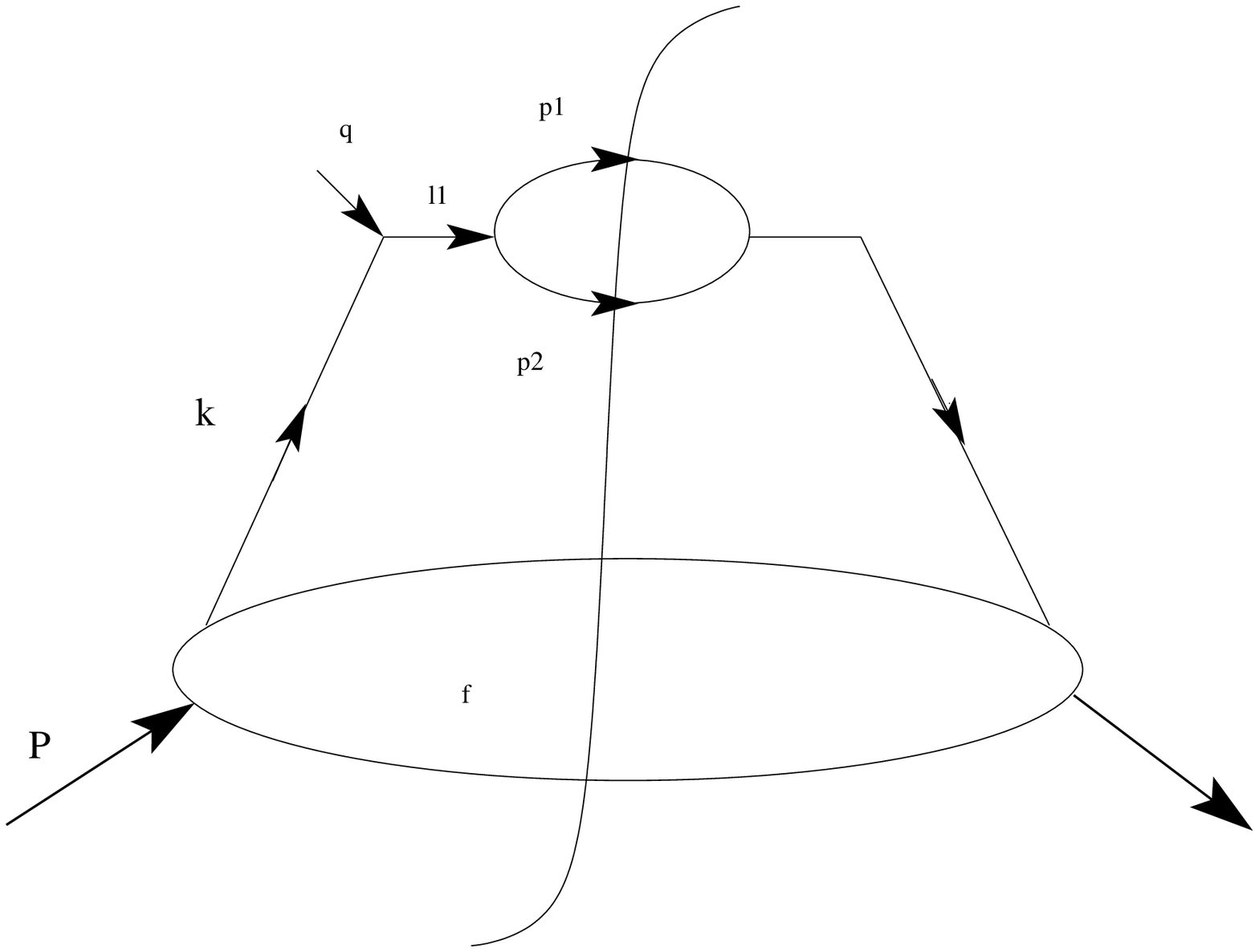}  
    \vspace{2mm}\\
        (a)&(b)&(c)
\vspace{7mm}\\
          \includegraphics[scale=.26]{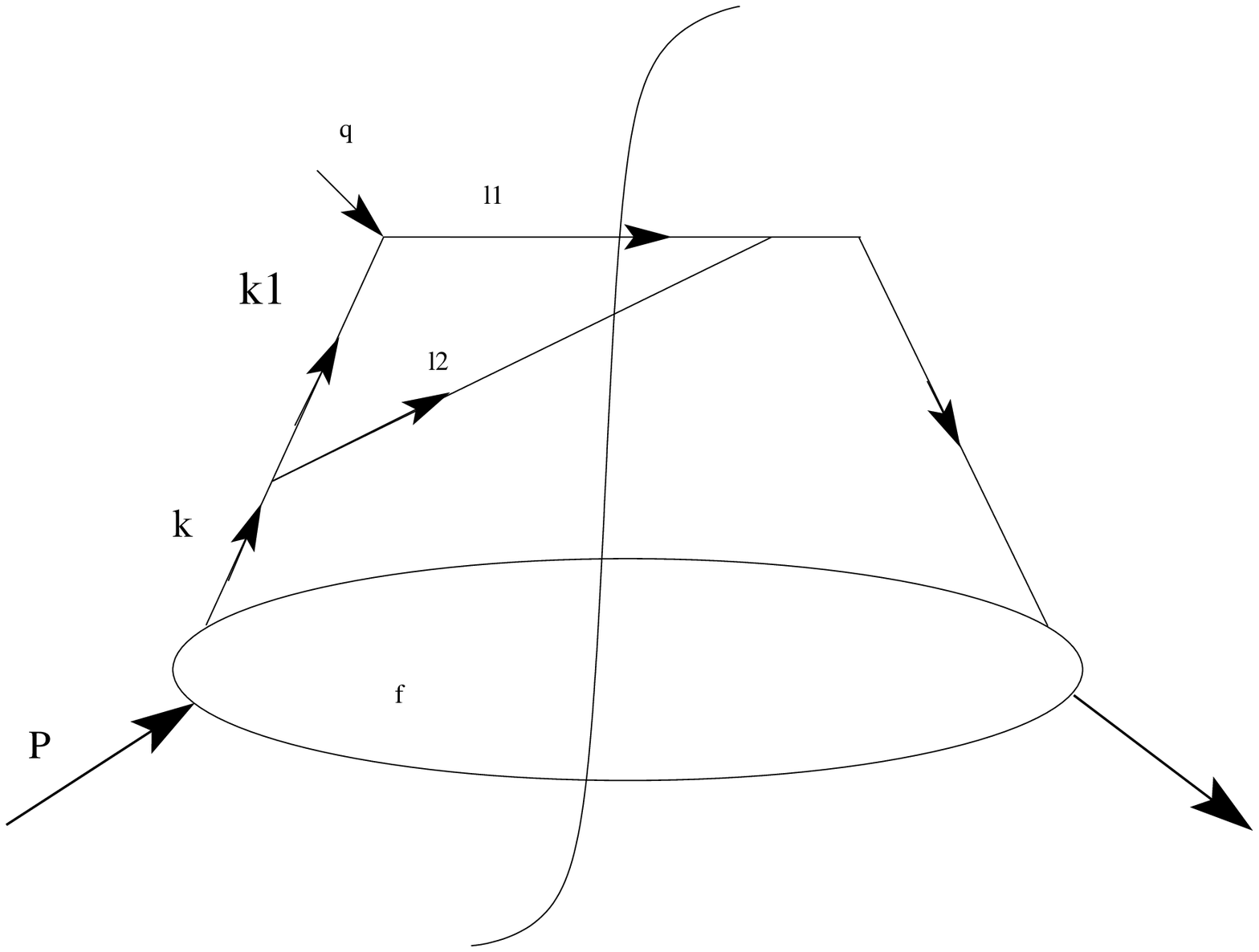}

         &
          \includegraphics[scale=.26]{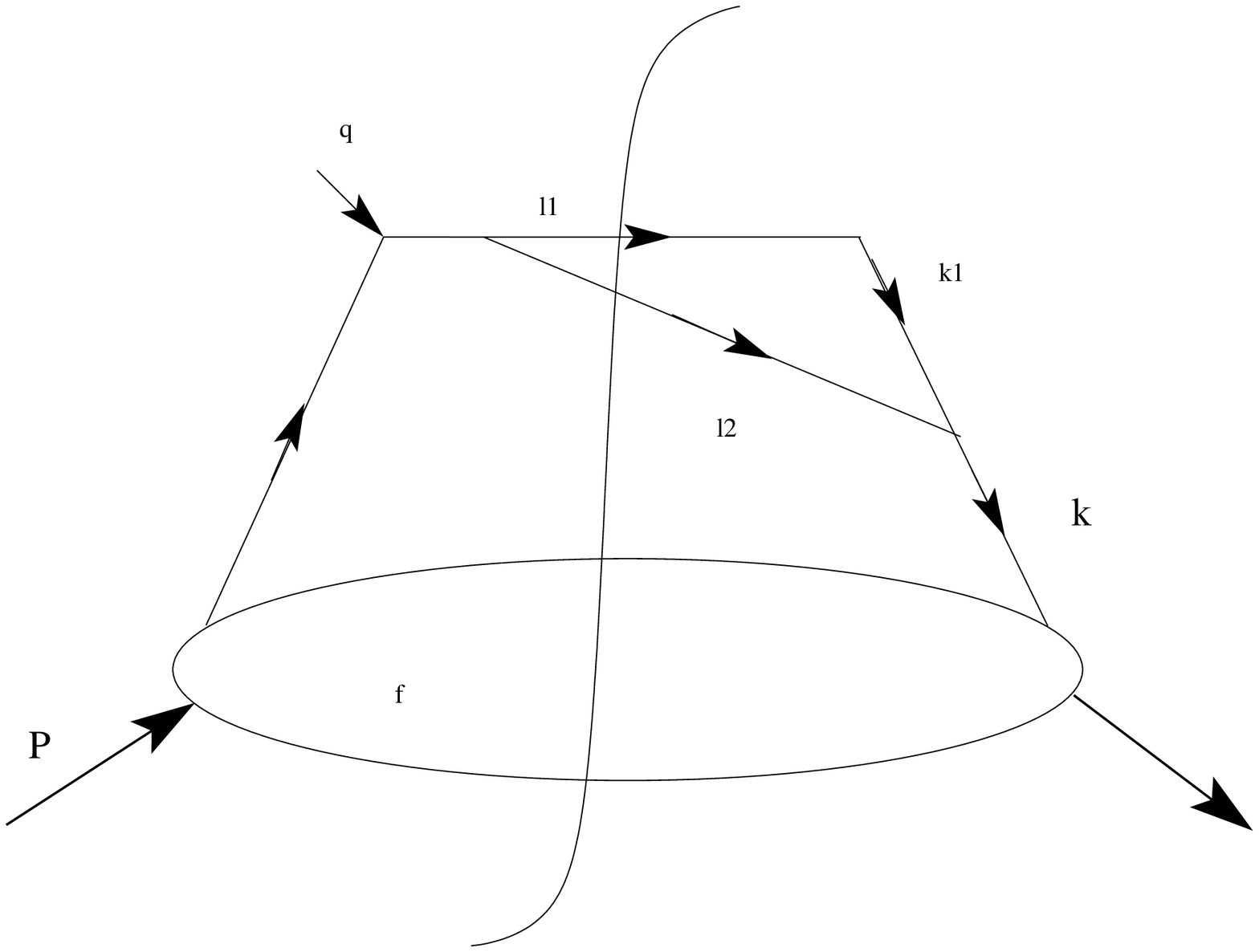} 
           &
          \includegraphics[scale=.26]{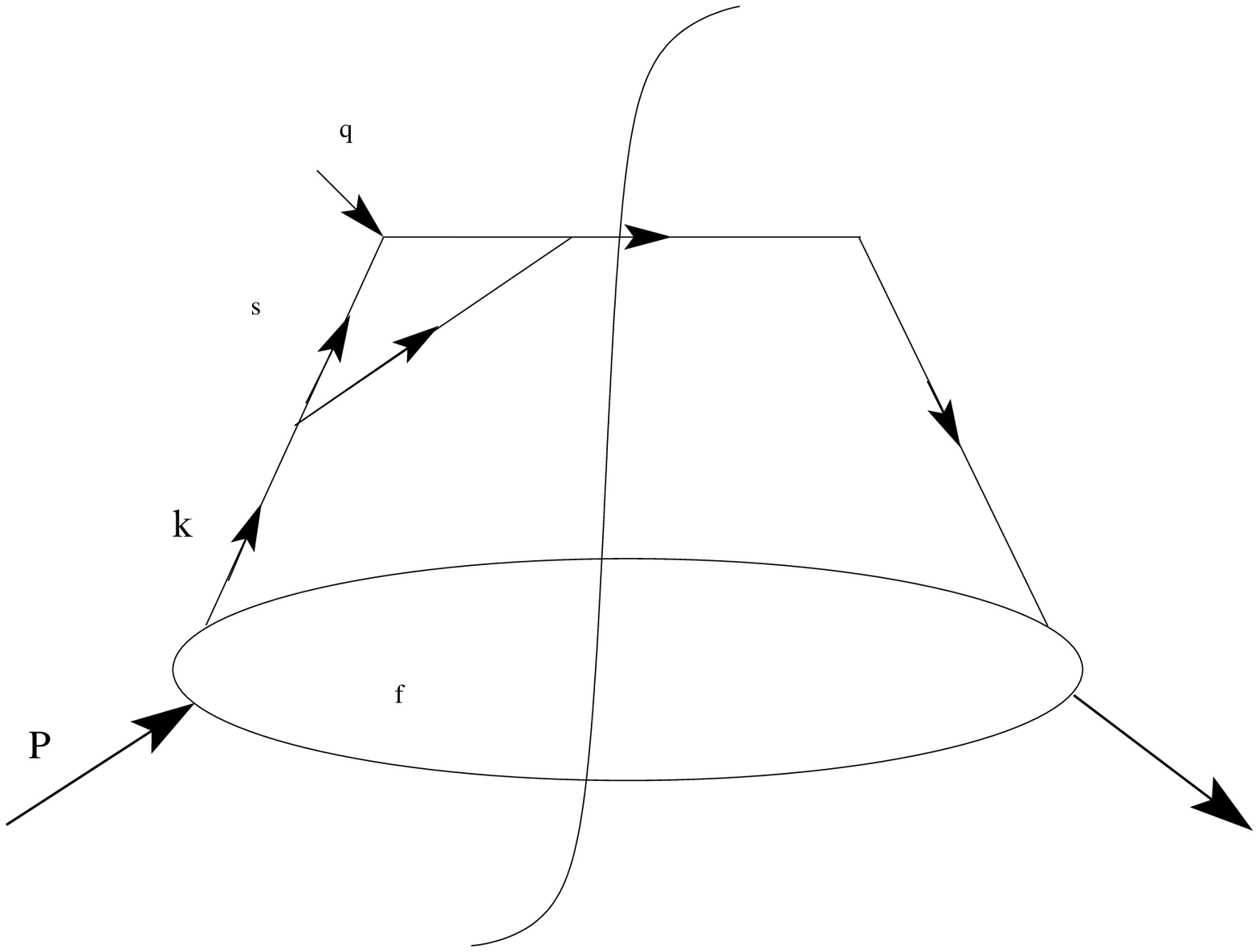}
    \vspace{2mm}\\
          (d)&(e) & (f) \vspace{2mm}\\
           &
          \includegraphics[scale=.26]{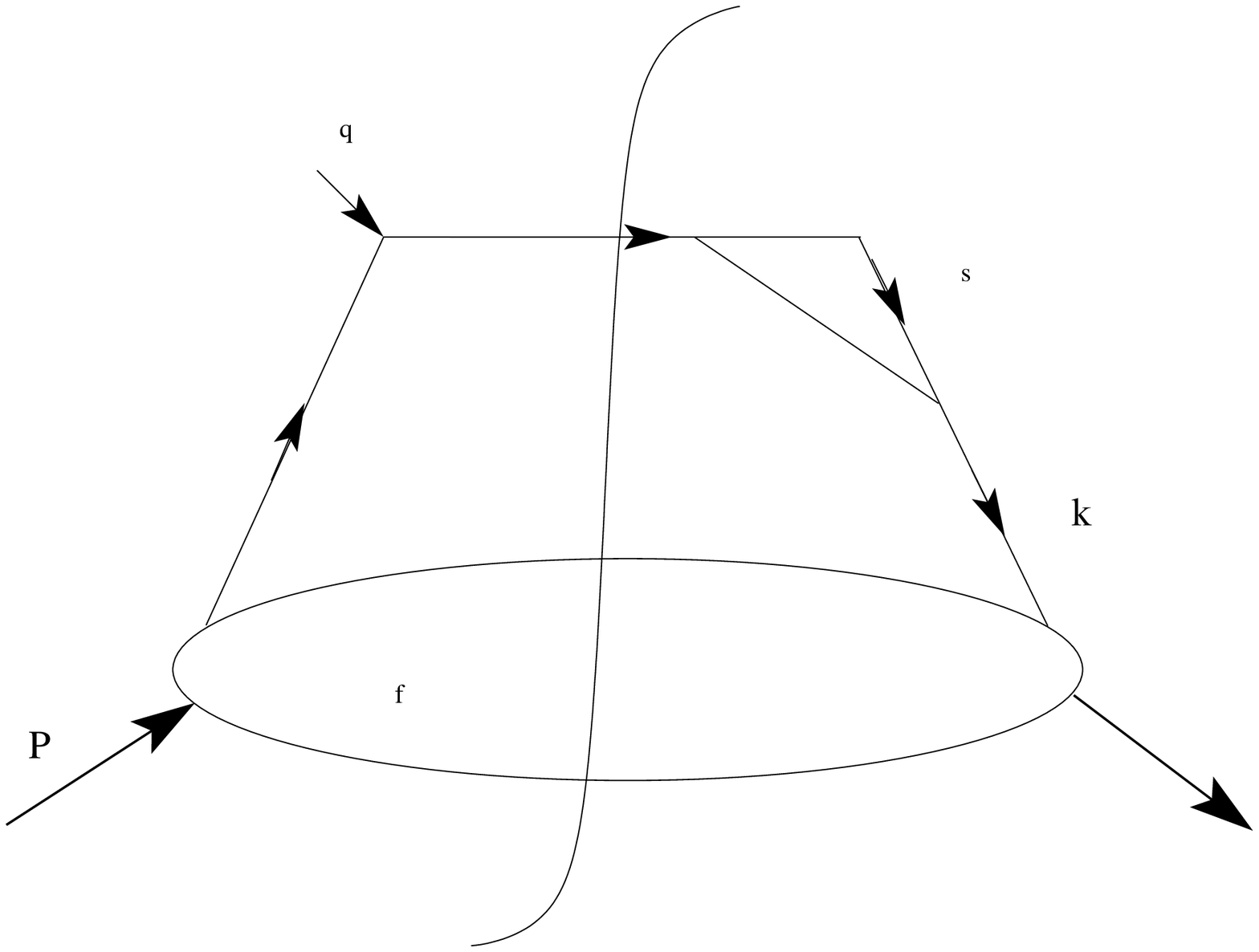} 
   \vspace{2mm}\\
         &(g)
\vspace{7mm}\\
        \end{tabular}
\caption{One-loop graphs for DIS.}
\label{fig:oneloop}
}

\begin{figure*}
\centering
\parbox{0.45\textwidth}{
\centering
    \psfrag{P}{\small $P$}
    \psfrag{f}{$\Phi(k,P)$}
    \psfrag{q}{$q$}
       \includegraphics[scale=.28]{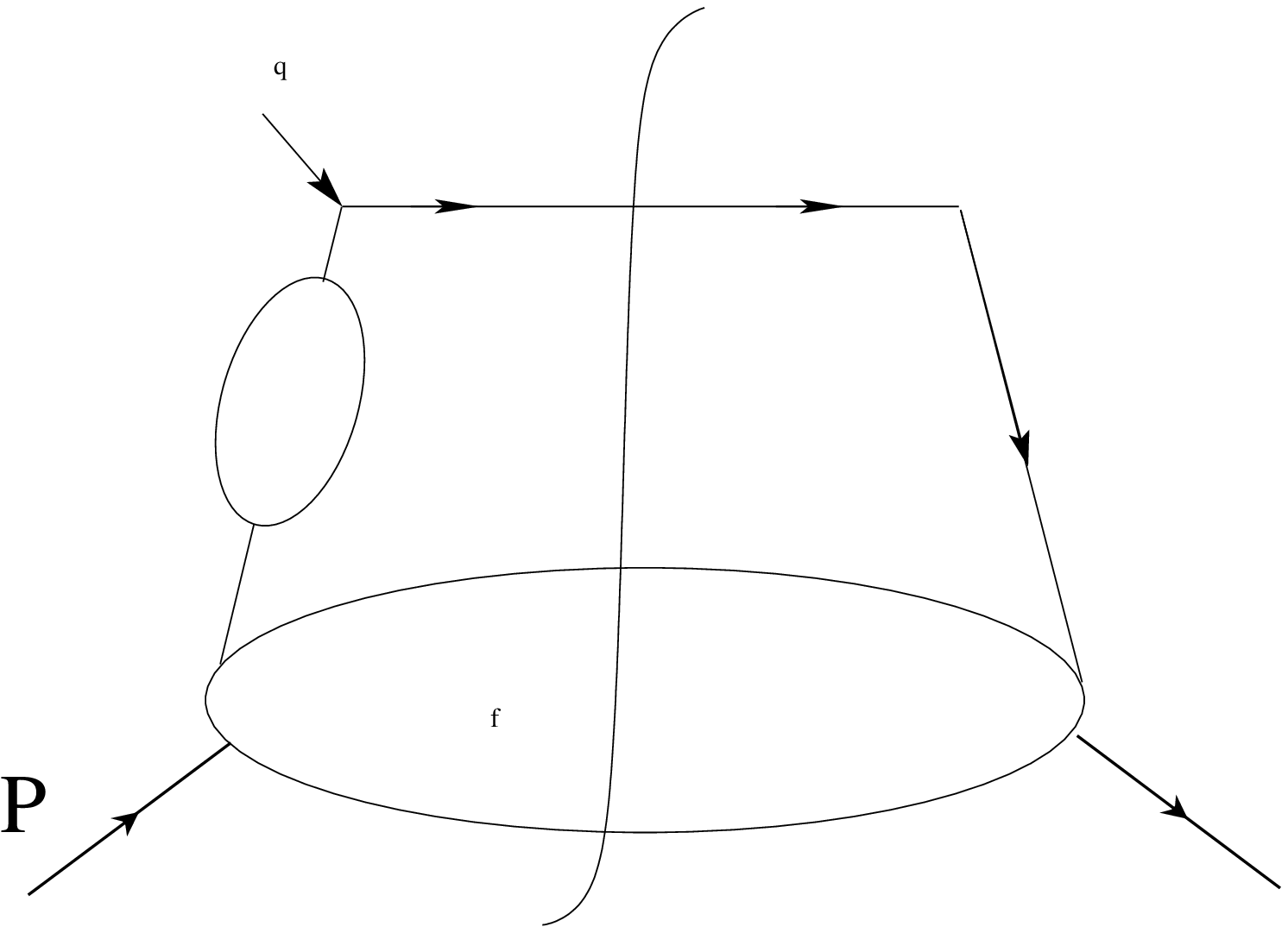}
}
\hfill
\parbox{0.45\textwidth}{
\centering
    \psfrag{P}{\small $P$}
    \psfrag{k}{\small $k$}
    \psfrag{f}{$\Phi(k,P)$}
    \psfrag{q}{$q$}
       \includegraphics[scale=.28]{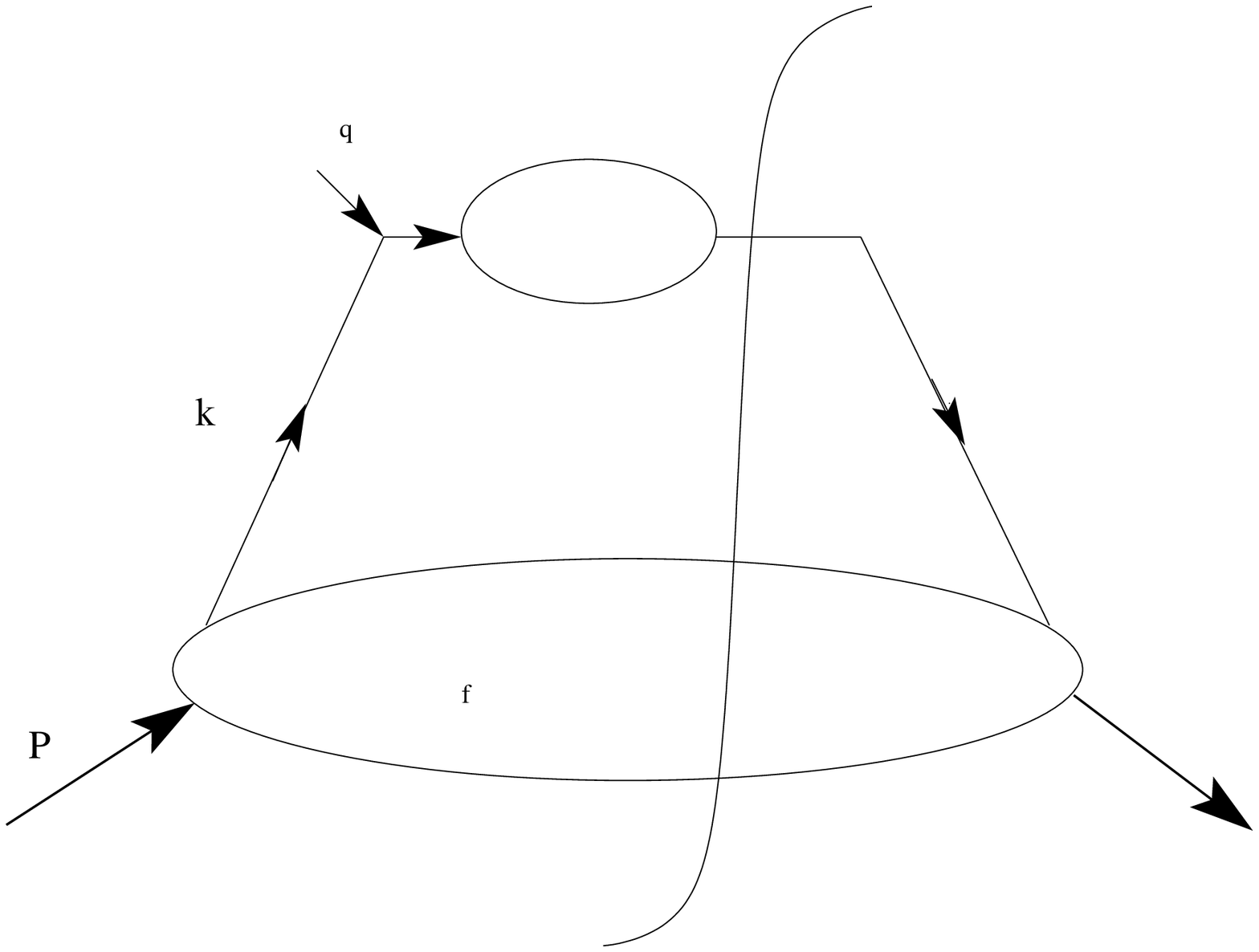}
}
\\
\parbox{0.45\textwidth}{
    \caption{A graph like this with an initial-state parton self energy is 
      not used.
    }
    \label{fig:initial.self.energy}
}
\hfill
\parbox{0.45\textwidth}{
    \caption{In a graph like this the final-state self energy
    contributes to the hadronization of the final-state parton, and
    such graphs are disregarded in computing the hard-scattering
    coefficient. 
    }
    \label{fig:final.self.energy}
}
\end{figure*}

\begin{figure*}
\centering
\parbox{0.45\textwidth}{
\centering
    \psfrag{k}{ $k$}
    \psfrag{P}{$P$}
    \psfrag{q}{ $q$}
    \psfrag{k1}{$k_1$}
    \psfrag{l1}{$\hat{l}_1$}
    \psfrag{l2}{$\hat{l}_2$}
    \psfrag{f}{$\Phi(k,P)$}
    \includegraphics[scale=.33]{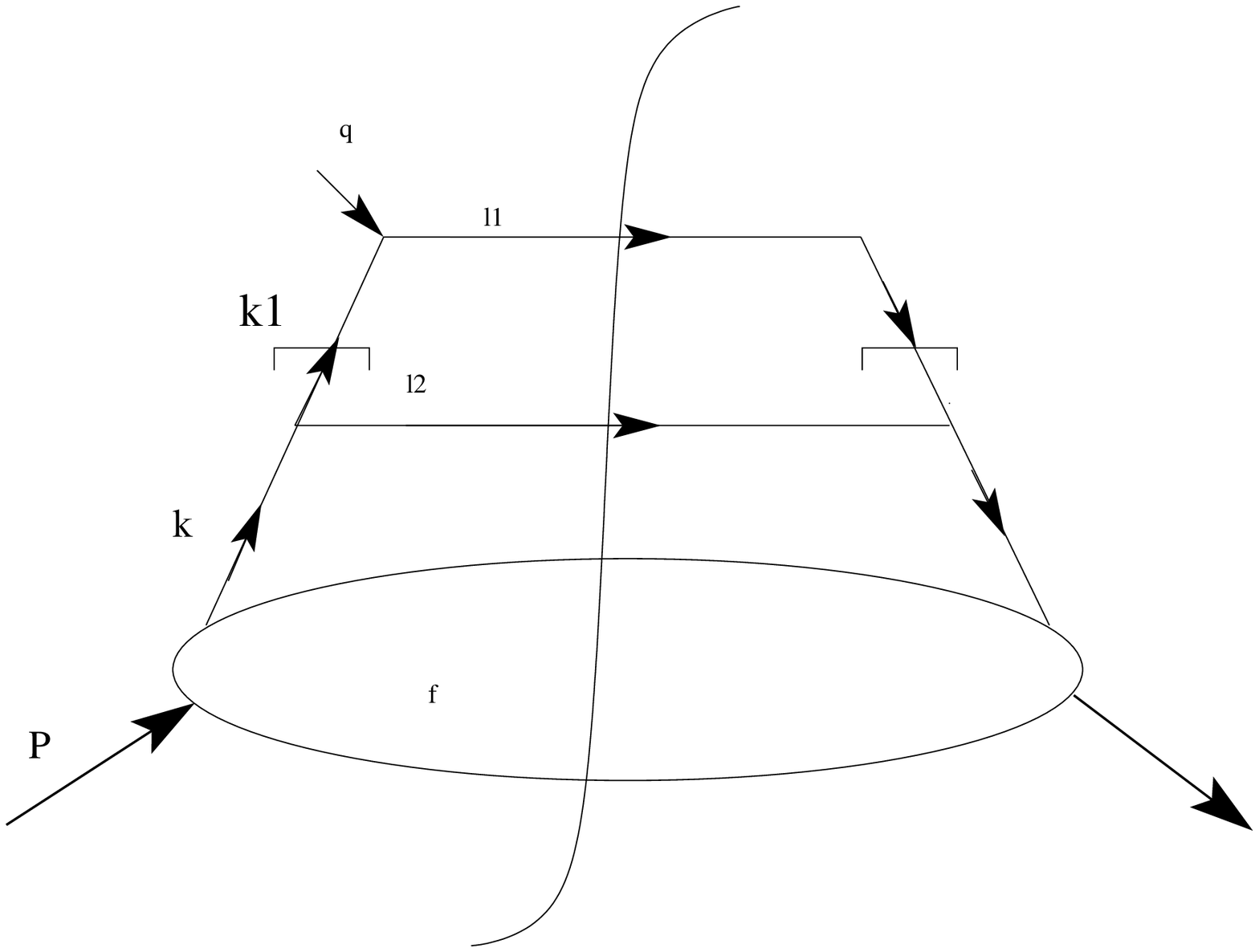}
}
\\
\parbox{0.45\textwidth}{
    \caption{Approximation for  region $R_1$ of Fig.\ 
    \protect\ref{fig:oneloop}(a). 
    The small square brackets indicate where the collinear
    approximation of Eq.\ (\protect\ref{eq:approx}) is made.
    }
    \label{fig:r1}
}
\end{figure*}

There are certain complications in the interpretation of these graphs.
Observe first that there is some apparent double counting.  For
example, the ladder graph (b) also implicitly appears in (a), with the
rung $l_2$ of (b) being inside the $\Phi$ subgraph of (a).  We
actually mean that we consider each term in Fig.\
\ref{fig:dis2} or \ref{fig:oneloop} not as a graph as such but as a
decomposition of a 
graph for the purposes of examining its behavior: a graph can have
multiple decompositions.  We will indeed sum over the different
decompositions, and the subtraction procedure in the hard
scattering will take care of the double counting.

Next, whereas we have used an explicit graphical factor $\Phi$ to
represent the connection to the target proton, we have omitted
corresponding fragmentation factors for the connection of the
final-state partons to the final-state hadrons.  A more thorough
presentation would have done this, but our focus in this paper is on
issues associated with initial-state partons.

Also, observe that self-energy corrections on the incoming parton line are
always included in the pCf.  Therefore we omit from our considerations
decompositions such as Fig.\ \ref{fig:initial.self.energy}; the
self-energy on the incoming line does \emph{not} contribute to our
calculation of the one-loop hard-scattering coefficients.

Similarly self-energy corrections on final-state parton lines, as in
Fig.\ \ref{fig:final.self.energy}, are always to be considered as part
of the hadronization, and these self-energies make no direct
contribution are therefore disregarded in the calculation of the
hard-scattering coefficients.  But observe that in the application to
an algorithm for a MC event generator, there are normalization factors
associated with final-state parton in the definition of the properly
normalization hard scattering \emph{cross section}
$d\hat{\sigma}_{N_u}$.  These are the factors of $Z$ in Eq.\
(\ref{eq:sigmah}), and their calculation, given in \cite{phi3}, does
indeed include a consideration of self-energy graphs.

\subsubsection{Zero-loop graph (a)}

Graph (a) has a tree graph coupled to the full pCf.  The hard
scattering contribution corresponds to the region where the incoming
parton line $k$ is collinear to the target.  Comparing the graph with
the general form of factorization, (\ref{eq:fact1a}) and
(\ref{eq:fact1b}), we see that the hard scattering coefficient is
\begin{equation}
  \hat{H}_1 = e^2 \, [1 + O(g^2)],
\end{equation}
with a corresponding cross section
\begin{equation}
  d\hat\sigma_1 = \frac{\pi}{q^-} \delta(\hat{k}^+-q^+) \,\,
             K e^2 \, [1 + O(g^2)]
          = \frac{\pi xP^+}{Q^2} \delta(\hat{k}^+-xP^+) \,\,
                K e^2 \, [1 + O(g^2)].
\end{equation}

\subsubsection{Graph (b)}

Now we treat the ladder graph, (b) of Fig.\ \ref{fig:oneloop}.  We
always suppose
$k$ is collinear to the target.  Then there are two leading regions. In
region $R_1$, $k_1$, $l_2$ and $k$ are almost collinear to the target,
with $m^2\sim |k^2| \leq |k_1^2| \ll Q^2$. In region $R_2$, $k_1$ and $l_2$
are at wide
angle, with\footnote{Here, for
  convenience of power counting, we choose $m$ to be the typical small
  scale. In general all we need is some scale $s$ which is much
  smaller than the hard scattering scale $Q$. That means, the
  virtuality of the external partons of the hard subgraph is of order
  $s^2$. If $s$ is much larger than the hadronization scale $\Lambda$ then
  we can recursively apply factorization to the jets generated by
  those external partons.} 
$m^2 \sim |k^2| \ll |k_1^2| \leq Q^2$.

First we must remember that, as noted earlier, graph (b) also appears
as a particular case of graph (a).  We constructed an approximation to
graph (a), and obtained the lowest-order hard scattering.  In terms of
graph (b), this approximation corresponds to the case that momentum
$k_1$ is almost collinear to the target, i.e., it is a good
approximation in region $R_1$.  Therefore our analysis of the regions
for (b) will be designed to obtain an appropriately subtracted hard
scattering corresponding to the other region.

This is done in terms of the approximation operations for the two
regions of graph (b), which correspond to different projections of the
parton momenta on massless momenta.  The projection for region $R_1$
we denote by\footnote{
   Notice that no approximation is made on $k$ or $l_2$ in this
   region; these momenta are considered as part of the pCf.
}
($\hat{k}_1$, $\hat{l}_1$, $l_2$, $k$) with Jacobian
$\Delta_1$, and the projection for region $R_2$ we denote by
($\tilde{k}_1$, $\tilde{l}_1$, $\tilde{l}_2$, $\tilde{k}$) with
Jacobian $\Delta_2$.  The values are:
\begin{eqnarray}
\label{eq:proj1}
  \hat{k}_1 &=& \left( -q^+, 0, \T{0} \right),
\nonumber\\
  \hat{l}_1 &=& \left( 0, q^-, \T{0} \right),
\\
\label{eq:delta1}
   \Delta_1 &=& 1+k_1^-/q^- = 1+2xP^+k_1^-/Q^2,
\end{eqnarray}
and
\begin{eqnarray}
\label{eq:proj2}
  \tilde{l}_1 &=& \left( \frac{ (\T{l_1}-\T{l_2})^2 }{ 4(q^-+l_1^--l_2^-) }, \,
                        \frac{ q^-+l_1^--l_2^- }{ 2 }, \,
                        \frac{ \T{l_1}-\T{l_2}}{ 2 }
                 \right),
\nonumber\\
  \tilde{l}_2 &=& \left( \frac{ (\T{l_1}-\T{l_2})^2 }{ 4(q^-+l_2^--l_1^-) }, \,
                        \frac{ q^-+l_2^--l_1^- }{ 2 }, \,
                        \frac{ \T{l_2}-\T{l_1}}{ 2 }
                 \right),
\nonumber\\
  \tilde{k}_1 &=& \left( -q^+ + \frac{ (\T{l_1}-\T{l_2})^2 }{ 4(q^-+l_1^--l_2^-) }, \,
                        \frac{ - q^-+l_1^--l_2^- }{ 2 }, \,
                        \frac{ \T{l_1}-\T{l_2}}{ 2 }
                 \right),
\nonumber\\
  \tilde{k}  &=& \left( -q^+ + \frac{ (\T{l_1}-\T{l_2})^2 }
                                   { 2[ (q^-)^2 - (l_1^--l_2^-)^2] }, \,
                        0, \,
                        \T{0}
                 \right),
\\
\label{eq:delta2}
   \Delta_2 &=& \frac{ l_1^-l_2^- }{ (l_1^--k^-/2) (l_2^--k^-/2) }
      = \frac{ 4l_1^-l_2^- }{ [ (q^-)^2 - (l_1^--l_2^-)^2] } .
\end{eqnarray}

Before approximation, graph (b) is represented by:
\begin{eqnarray}
   \int \Gamma_b &=&\int \frac{ d^6k }{ (2\pi)^6 } \Phi(k,P) 
         \int  \frac{ d^6k_1 }{ (2\pi)^6 } \frac{ e^2g^2 }{ (k_1^2-m^2)^2 }
        \int \frac{d^5\vec{l}_1}{(2\pi)^5 2E_1} (2\pi)^6 \delta^{(6)}(l_1-k_1-q) 
\nonumber \\
        && \times \int \frac{d^5\vec{l}_2}{(2\pi)^5 2E_2} (2\pi)^6 
        \delta^{(6)}(l_2-k+k_1). 
\end{eqnarray}
Our goal is to apply the general decomposition by regions that we gave
in Eqs.\ (\ref{eq:regions}), (\ref{eq:fac-app}) with our chosen
definition of the approximation operator $T_R$.  We will construct
terms $\Gamma_{b1}=C_{R1}(\Gamma_b)$ and $\Gamma_{b2}=C_{R2}(\Gamma_b)$ and will
explicitly check that the sum is a good approximation to the original
graph:
\begin{equation}
\Gamma_b = \Gamma_{b1}+ \Gamma_{b2} + O(|\Gamma_b|m^2/Q^2)
\end{equation} 
everywhere that $k$ is collinear to $P$.

\paragraph{Region $R_1$.} In this region, $|k_1^2|$ is $O(m^2)$, so $l_2$ and 
$k$ are almost parallel. The corresponding approximation, symbolized in
Fig.\ \ref{fig:r1}, is 
\begin{equation}
\label{eq:ladderR1}
        \int \Gamma_{b1}
       = \int T_{R1} \Gamma_b 
       = \int \frac{d\hat{k}^+}{2\pi}
          \int \frac{d^5k}{(2\pi)^5}\Phi^{(1)}(k_1,P) \,\,
          dL(\hat{l}_1; q+\hat{k}_1) \hat{H}_1(\hat{l}_1, \hat{k}_1; q, m=0),  
\end{equation}
with
\begin{equation}
\Phi^{(1)}(k_1,P) = \int \frac{d^6k}{(2\pi)^6} \Phi(k,P)\,\,
           dL(l_2; k-k_1) \frac{e^2g^2}{(k_1^2-m^2)^2}
\end{equation}
representing a one-rung iteration of the pCf.  The approximated
momenta are defined by Eq.\ (\ref{eq:proj1}).  The result in Eq.\ 
(\ref{eq:ladderR1}) gives of course just an example of a lowest order
hard scattering, and we already know that the contribution to the hard
scattering factor is $\hat{H}_1(\hat{l}_1, \hat{k}_1; q, m=0) = e^2$.

When constructing the subtractions for region $R_2$, we need the formula
for $\Gamma_{b1}$ in terms of the original phase space integration:
\begin{equation}
        \int \Gamma_{b1}
      = \int \frac{d^6k}{(2\pi)^6} \Phi(k,P)
        \int \frac{d^6k_1}{(2\pi)^6} \,dL(l_1; k_1+q)  \,\,dL(l_2; k-k_1)
                \,\,\Delta_1 \frac{e^2g^2}{(k_1^2-m^2)^2}.
\end{equation}
Also we need to verify that $\Gamma_{b1}$ is indeed a good approximation to $\Gamma_b$
in region $R_1$:
\begin{eqnarray}
\label{eq:esr1}
  \int (\Gamma_b - \Gamma_{b1}) &=& \int \frac{d^6k}{(2\pi)^6} \Phi(k,P)
        \int \frac{d^6k_1}{(2\pi)^6} \,dL(l_1; k_1+q)  \,\,dL(l_2; k-k_1)
\nonumber \\
    && \times \frac{e^2g^2}{(k_1^2-m^2)^2}
        (1-\Delta_1) .
\end{eqnarray}
Since $1-\Delta_1 = k_1^-/q^-$, which is small in $R_1$, the desired
suppression follows.

\paragraph{Region $R_2$.} In this region, both of the final-state
partons $l_1$ and $l_2$ are at wide
angle.  The approximation for the whole graph in this region is
\begin{eqnarray}
        \int T_{R2} \Gamma_b &=& 
         \int \frac{ d\tilde{k}^+dk^- d^4\T{k} }{ (2\pi)^6 } \Phi(k,P) \,
        dL(\tilde{l}_1, \tilde{l}_2; q+\tilde{k}; m=0) 
\nonumber \\
       &&\times  \int \frac{d^6\tilde{k}_1}{(2\pi)^6} \delta^{(6)}
        (\tilde{l}_2 - (\tilde{k}_1-\tilde{k})) H_2. 
\end{eqnarray}
Here $H_2 = e^2g^2/(\tilde{k}_1^2)^2$,
while $\tilde{k}$, $\tilde{k}_1$, $\tilde{l}_1$, and $\tilde{l}_2$ are
given by (\ref{eq:proj2}).

The hard scattering factor $H_2$
has a collinear singularity when $\tilde{l}_2$ and
$\tilde{k}$ become parallel. However, this is in the smaller
region $R_1$, where
$\Gamma_{b1}$ is a good approximation, and for which we are required to
apply a subtraction, according to the general formula (\ref{eq:fac-app}).
Therefore we define $\Gamma_{b2}$ to be the approximation of the original
graph $\Gamma_b$ {\it minus} the approximation in $R_1$, i.e., $\Gamma_{b1}$.
 Then the singularity in the collinear region will be canceled
point-by-point in the integration space:
\begin{eqnarray}
\label{eq:Gam.1b}
   \int \Gamma_{b2}
&=& \int V(-\tilde{k}_1^2/m^2)
        \,T_{R2} (\Gamma_b -\Gamma_{b1})
\nonumber\\
&=&
        \int \frac{d^6k}{(2\pi)^6} \Phi(k,P)
        \int \frac{d^6k_1}{(2\pi)^6} \int \frac{d^5\vec{l}_1}{(2\pi)^5 2E_1}
        (2\pi)^6 \delta^{(6)}(l_1 -q-k_1) 
\\
        && \times \int \frac{d^5\vec{l}_2}{(2\pi)^5 2E_2}
             (2\pi)^6 \delta^{(6)}(l_2-k+k_1) \,
             V(-\tilde{k}_1^2/m^2) \,
            \frac{e^2g^2}{(\tilde{k}_1^2)\strut^2} 
           (1-\Delta'_{1b}) \Delta_2.
\nonumber
\end{eqnarray}
Here $\Delta'_{1b}$ is the Jacobian in the subtraction term with the momenta
$k$, etc. replaced by the approximated momenta $\tilde{k}_1$, etc.,\
that are appropriate for  region $R_2$, i.e.,
\begin{eqnarray}
\label{eq:del-1b'}
     \Delta'_{1b} &=& \frac{\tilde{l}_1^-}{\tilde{l}_1^- - \tilde{k_1}^-}
          = \frac{l_1^- - k^-/2}{l_1^- -k^-/2 +\tilde{l_2}^-}
          = \frac{l_1^- - k^-/2}{l_1^- - k_1^-} 
\nonumber\\
   &=& \Delta_1 -\frac{k^-}{2(l_1^--k_1^-)}
    = \Delta_1 \left[ 1+O\!\left(\frac{m^2}{Q^2}\right) \right].
\end{eqnarray}
The Jacobian $\Delta_2$ is given by Eq.\ (\ref{eq:jac}) with $N_u = 2$.
The collinear veto factor $V(\tilde{k}_1/m)$, defined in  Eq.\ (\ref{eq:veto}),
 is used to remove a spurious \emph{nonleading}
collinear divergence in the differential cross section due to the factor
 $H_2= e^2g^2/(\tilde{k}_1^2)^2$.  Since
$\Gamma_{b2}$ is designed to handle momentum configurations far from the 
collinear region, the precise functional form of the veto factor in the
collinear region is irrelevant to physics. Any other function which is zero
when $|\tilde{k}_1^2|  \ll m^2$ and  unity when $|\tilde{k}_1^2| \gg m^2$
would be equally good.

It follows from Eq.\ (\ref{eq:Gam.1b}) that the contribution of the
one-rung ladder to the hard scattering is
\begin{equation}
\label{eq:H.b}
    V(-\tilde{k}_1^2/m^2) \,
    \frac{e^2g^2}{(\tilde{k}_1^2)\strut^2} 
    (1-\Delta'_{1b}).
\end{equation}

\paragraph{Validity of approximations}
We can check that $\Gamma_{b2}$ is indeed power suppressed in region $R_1$.
\begin{equation}
  \Gamma_{b2} = V T_{R2} ( \Gamma_b - \Gamma_{b1}) = O\left(|\Gamma_b-\Gamma_{b1}|
   \frac{(k_1^2)^2}{(\tilde{k}_1^2)^2}\right) = O\left(|\Gamma_b| 
   \frac{|k_1^2|}{Q^2}\right) 
\stackrel{R_1}{=}
    O\left(|\Gamma_b| 
   \frac{m^2}{Q^2}\right) ,
\end{equation}
where we have used Eq.\ (\ref{eq:esr1}) to estimate the size of $\Gamma_b-\Gamma_{b1}$.
In region $R_2$, where $|k^2|$ is the typical scale that we ignored and 
$|k_1^2|$ is the typical hard scale, we can see the following is true.
\begin{equation}
    \Gamma_{b2}-(\Gamma_b-\Gamma_{b1}) 
= O \left( \left | \frac{k^2}{k_1^2}\right| |  \Gamma_b - \Gamma_{b1}|\right)
 = O \left( \frac{m^2}{Q^2} |\Gamma_b|\right).
\end{equation}

Combining the error estimates in different regions we can see that $\Gamma_{b1}$ and 
$\Gamma_{b2}$ indeed satisfy our requirement that $\Gamma_b = \Gamma_{b1}+ \Gamma_{b2} 
+ O(|\Gamma_b|m^2/Q^2)$ in all regions.

\subsubsection{Graph (c)}
Next we treat graph (c) of Fig.\ \ref{fig:oneloop}.  The treatment is
essentially the same as that of graph (b), except that the subtraction
is about 
final-state fragmentation.  The graph has two leading regions. For its
smallest region, for which we use the notation
$R_{c1}$, $p_1$ and $p_2$ are almost parallel. In region $R_{c2}$, $p_1$ and
$p_2$ are at wide angle. Since final state
subtractions have been treated in detail in \cite{phi3}, we will give
the result directly without details. The major difference is in the
projections to massless momenta in DIS and $e^+e^-$.  Again, we use
symbols like $\hat{l}_1$ for approximated momenta in region $R_{c1}$ and 
symbols like $\tilde{l}_1$ for those in region $R_{c2}$.

The graph can be represented as 
\begin{eqnarray}
  \int \Gamma_c &=& \int \frac{d^6k}{(2\pi)^6} \Phi(k,P) \int \frac{d^5 \vec{p}_1}{(2\pi)^5 2E_1}
      \int \frac{d^5 \vec{p}_2}{(2\pi)^5 2E_2}
     (2\pi)^6 \delta^{(6)}\boldsymbol((q+k)-(p_1+p_2)\boldsymbol) \nonumber \\
      && \times \frac{e^2g^2}{(q+k)^2-m^2}.
\end{eqnarray}

\paragraph{Region $R_{c1}$.}
In region $R_{c1}$ of graph (c), $\Gamma_c$ is approximated as 
\begin{eqnarray}
   \int \Gamma_{c1} &=& \int T_{R_{c1}} \Gamma_{c} 
\nonumber\\
   &=& \int \frac{d\hat{k}^+}{2\pi}
      \int \frac{dk^- d^4 \T{k}}{(2\pi)^5} \Phi(k,P) 
      \int \frac{d^5\hat{\vec{l}}_1}{(2\pi)^5 2|\hat{\vec{l}}_1|}
      (2\pi)^6 \delta^{(6)}(\hat{l}_1 -q-\hat{k}) 
\nonumber \\
      &&\times \int \frac{d M_1^2}{2\pi} \int \frac{d^5\vec{p}_1}{(2\pi)^5 2E_1}
      \int \frac{d^5\vec{p}_2}{(2\pi)^5 2E_2}(2\pi)^6 \delta^{(6)}(l_1-p_1-p_2) 
      \frac{e^2g^2\Theta(M_1^2/\mu_J^2)}{[(p_1+p_2)^2-m^2]^2}.
\nonumber \\
     &=& \int  \frac{d^6k}{(2\pi)^6} \Phi(k,P) \int \frac{d^5 \vec{p}_1}{(2\pi)^5 2E_1}
      \int \frac{d^5 \vec{p}_2}{(2\pi)^5 2E_2}
      (2\pi)^6 \delta^{(6)}(q+k-p_1-p_2)
\nonumber \\
      && \times \frac{e^2g^2\Theta(M_1^2/\mu_J^2) \Delta_{c1}  }{(q+k)^2-m^2},
\end{eqnarray}
where the Jacobian is $\Delta_{c1} = l_1^- /(l_1^- - k^-)$ and the mass of
parton $l_1$ is $M_1^2=l_1^2=(p_1+p_2)^2$.

This value is in fact a full pCf times the lowest-order hard
scattering times an order $g^2$ approximation to final-state
showering.\footnote{Note that, whereas we drew graph (a) with an
   explicit factor for the initial-state interactions, we did not
   explicitly give it its corresponding final-state factor.
}
Thus it is to be found as a contribution to full
factorized cross section with the lowest-order hard scattering; there
is no new contribution to the hard scattering.

\paragraph{Region $R_{c2}$.} 
For region $R_{c2}$ of graph (c), we have the approximation
\begin{eqnarray}
   \int T_{R_{c2}}\Gamma_c &=& \int \frac{d\tilde{k}^+}{2\pi} \int\frac{dk^-d^4\T{k}}{(2\pi)^5}
        \Phi(k,P)
       \int \frac{d^5 \tilde{\vec{p}}_1}{(2\pi)^5 |\tilde{\vec{p}}_1|}
       \int \frac{d^5 \tilde{\vec{p}}_2}{(2\pi)^5 |\tilde{\vec{p}}_2|}
       (2\pi)^6 \delta^{(6)}(\tilde{p}_1 + \tilde{p}_2 - q - \tilde{k}) 
\nonumber \\
       && \times \frac{e^2g^2}{[(\tilde{p}_1 + \tilde{p}_2)^2]^2},
\end{eqnarray}
which is divergent in the collinear region $R_{c1}$. After subtraction we 
have
\begin{eqnarray}
 \int \Gamma_{c2} &=& \int V(\tilde{M}_1^2/m^2) \, T_{R_{c2}}(\Gamma_c - \Gamma_{c1}) 
\nonumber\\
   &=& \int 
     \frac{d^6k}{(2\pi)^6} \Phi(k,P) \int \frac{d^5 \vec{p}_1}{(2\pi)^5 2E_1}
      \int \frac{d^5 \vec{p}_2}{(2\pi)^5 2E_2}  
      (2\pi)^6 \delta^{(6)}(q+k-p_1-p_2) \,
\nonumber\\
      && \times \,\,
      V(\tilde{M}_1^2/m^2) 
      \frac{ e^2g^2 }{ [(\tilde{p}_1 + \tilde{p}_2)^2]^2 }
      [1-\Theta((\tilde{p}_1 + \tilde{p}_2)^2/\mu_J^2) \Delta_{1c}'] \,
      \Delta_{c2} ,
\end{eqnarray}
where $\Delta_{1c}'$ is the Jacobian calculated with massless momenta for region $R_2$
\begin{equation}
\label{eq:del-1c'}
  \Delta_{1c}' = \frac{\tilde{p}_1^- + \tilde{p}_2^-}{\tilde{p}_1^- + \tilde{p}_2^- -k^-} =
  \frac{p_1^-+p_2^--k^-}{p_1^-+p_2^--2k^-}=\Delta_1(1+O(k^-/l_1^-))=\Delta_1(1+O(m^2/Q^2)).
\end{equation}
$\Delta_{c2}$ is defined using Eq.\ (\ref{eq:jac}) with $N_u=2$.
It is easy to check that $\Gamma_c=\Gamma_{c1}+\Gamma_{c2}+O(m^2/Q^2)$ in all
regions.

It follows that there is the following contribution to the hard scattering
$\hat{H}_2$:
\begin{equation}
  \label{eq:H.c}
      V(\tilde{M}_1^2/m^2) 
      \frac{ e^2g^2 }{ [(\tilde{p}_1 + \tilde{p}_2)^2]^2 }
      \left[1-\Theta\left( \frac{(\tilde{p}_1 + \tilde{p}_2)^2}{\mu_J^2} \right)
            \Delta_{1c}'
      \right].
\end{equation}

\subsubsection{Graphs (d) and (e)}
Graphs (d) and (e) of Fig.\ \ref{fig:oneloop} are hermitian conjugates
of each other.  They each have only one leading region, where the
two final-state particles are at wide angle.  We have
\begin{eqnarray}
    \int \Gamma_{d}&=&\int \frac{d^6k}{(2\pi)^6} \Phi(k,P) 
    \int \frac{d^6k_1}{(2\pi)^6}\int \frac{d^5 \vec{l}_1}{(2\pi)^5 2E_1}
    \int \frac{d^5 \vec{l}_2}{(2\pi)^5 2E_2}\,(2\pi)^6 \delta^{(6)}(l_1-q-k_1)
\nonumber \\
    && \times (2\pi)^6 \delta^{(6)}(l_2-k+k_1)
    \frac{ e^2g^2 }{ (k_1^2-m^2) \, [(q+k)^2-m^2] } 
\\
    \int \Gamma_{e} &=& {\rm H.c.} \int \Gamma_d = \int \Gamma_d,
\\
    \int \Gamma_{d1} &=& \int T_R \Gamma_d
  = \int \frac{d^6k}{(2\pi)^6} \Phi(k,P) 
   \int \frac{d^5 \vec{l}_1}{(2\pi)^5 2E_1} \int \frac{d^5 \vec{l}_2}{(2\pi)^5 2E_2}
\nonumber \\
   && \times\,(2\pi)^6  \delta^{(6)}(l_1+l_2-q-k)
      \frac{ e^2g^2\Delta_2 }{ (q+\hat{k})^2(\hat{l_2}-\hat{k})^2 },
\end{eqnarray}
where $\Delta_2$ and the approximated momenta ($\hat{k}$, $\hat{l_2}$, \dots) are defined
using Eq.\ (\ref{eq:jac}) and Eq.\ (\ref{eq:proj}) with $N_u=2$.

The contribution to the hard scattering of these two graphs is
\begin{equation}
  \label{eq:H.de}
  2
  \frac{ e^2g^2 }{ (q+\hat{k})^2(\hat{l_2}-\hat{k})^2 }.
\end{equation}

\subsubsection{Virtual corrections: graphs (f) and (g)}
Graph (f) and (g) in Fig.\ \ref{fig:oneloop} provide vertex
corrections to one-parton production.  They have only one leading
region, where the loop momentum is hard.  There is also a UV
divergence in the loop integral which we renormalize using the
\MSbar-scheme, so that
\begin{eqnarray}
  \int \Gamma_f &=& \int \frac{d^6k}{(2\pi)^6} \Phi(k,P) \int 
     \frac{d^5 \vec{l}_1}{(2\pi)^5 2E_1}\,(2\pi)^6 \delta^{(6)}(l_1-q-k)
\\
     && \times \int \frac{d^6k_1}{(2\pi)^6} \frac{e^2g^2}{(k_1^2-m^2) \,
       [(q+k_1)^2-m^2] \, [(k-k_1)^2-m^2] }  + \mbox{UV counterterm},
\nonumber \\
  \int \Gamma_g &=& {\rm H.c.}\int \Gamma_g \\
  \int \Gamma_{f1}&=& \int T_R \Gamma_f = \int \frac{d^6k}{(2\pi)^6} \Phi(k,P) \int 
     \frac{d^5 \vec{l}_1}{(2\pi)^5 2E_1}\,
    (2\pi)^6 \delta^{(6)}(l_1-q-k) \Delta_1
\nonumber \\
     && \times \int \frac{d^6k_1}{(2\pi)^6} \frac{e^2g^2}{k_1^2(q+k_1)^2(\hat{k}-k_1)^2}
     +\mbox{\MSbar \,\, counterterm}, 
\nonumber \\
     &=&\int \frac{d^6k}{(2\pi)^6} \Phi(k,P) \int 
     \frac{d^5 \vec{l}_1}{(2\pi)^5 E_1}\,(2\pi)^6 \delta^{(6)}(l_1-(q+k)) 
 \nonumber      \\     
 &&\times \Delta_1
     \frac{e^2g^2}{128\pi^3} \left( \ln \frac{Q^2}{\mu^2} - 3 \right).
\end{eqnarray}
Note that the vertex correction depends on the renormalization scale
$\mu$; this just shows that the ``current'' $j = 1/2 [\phi^2]$ has an
anomalous dimension. The renormalization group of the current can be
found in \cite{phi3}.  The anomalous dimension is canceled by the
anomalous dimension of $e^2$.

\subsubsection{Total}
Combining the $O(g^2)$ hard-scattering coefficients from graphs (b) to
(e) in Fig.\ \ref{fig:oneloop}, we then have the
hard-scattering coefficient for $\gamma^* + \mbox{parton} \to p_1 + p_2$:
\begin{eqnarray}
   \hat{H}_2
&=&  V\boldsymbol( -(\tilde{k}-\tilde{p}_2)^2 / m^2 \boldsymbol) 
        \frac{e^2g^2}{ [(\tilde{k}-\tilde{p}_2)^2]^2 } 
        (1-\Delta'_{1b})
        + \frac{2e^2g^2}{(q+\tilde{k})^2(\tilde{p}_2-\tilde{k})^2}
\nonumber \\
   &&+\,V\boldsymbol( (\tilde{p}_1+\tilde{p}_2)^2 / m^2 \boldsymbol) 
      \frac{ e^2g^2 }{ [(\tilde{p}_1 + \tilde{p}_2)^2]^2 }\,
      [1-\Theta((\tilde{p}_1 + \tilde{p}_2)^2/\mu_J^2) \Delta_{1c}'] + O(e^2g^4),
\end{eqnarray}
with $\Delta_{1b}'$ and $\Delta_{1c}'$ defined in Eq.\,(\ref{eq:del-1b'}) and
(\ref{eq:del-1c'}), respectively, and $V$ being the veto function that
cuts out the singularities. The approximated momenta $\tilde{p}_1$, 
{\it etc}. are defined in Sec.\,\ref{sec:proj}  with $N_u=2$.

From graphs (a), (f) and (g) we find the one-parton term with one-loop
corrections:
\begin{eqnarray}
  \hat{H}_1 = e^2 \left[
               1 
               + \frac{g^2}{64\pi^3}
                 \left( \ln \frac{Q^2}{\mu^2} - 3 \right)
               + O(g^4)
             \right].
\end{eqnarray}

%----------------------------------------
\subsection{Parton correlation function}

In our model theory, the current for the ``photon'' is just a scalar
vertex with two parton lines.  Graphs for the parton correlation
function are the same as for the structure function multiplied by:
\begin{itemize}
\item Two \emph{full} parton propagators at momentum $k$, i.e., a
  factor $1/(k^2)\strut^2$ at leading order.
\item A factor of $g^2$ instead of the factor $e^2$ associated with the 
  vertex for $e^2[\phi^2]/2$.
\item A factor $-2\pi/k^2$ to remove the prefactor $Q^2/(2\pi)$
  in our definition of
  the structure function---see Eq.\ (\ref{eq:parton.model.graph}). 
\end{itemize}

%==========================================================
\section{Conclusions and future work}
\label{sec:cl}

We have shown that in a full treatment of a Monte-Carlo event
generator the derivation of 
standard factorization is not sufficient.  The physics in standard
factorization is indeed unchanged.  What is at issue is the
validity of the identification, for example, of the momenta of the
external partons of a lowest-order hard scattering with the naive
parton model values, Eq.\ (\ref{eq:PM.momenta}).  It is not just that
showering of the parton $k$ in Fig.\ \ref{fig:handbag} gives it
nonzero values $k^-$ and $\T{k}$, for that is implicit already in a
correct definition of a parton density.  The problem that forces us to
change views is that $k^+$ is changed by the showering of the other
parton.  We have shown how these issues can be handled by the use of a
new factorization theorem that uses unintegrated parton correlation
functions, and in a model theory we have derived an appropriate
theorem.

In the new factorization theorem, we consider the structure function
differential in the full final state, or more conveniently we consider
the structure function with an arbitrary weight function applied to
the final state. The form of the factorization enables a systematic
first level of decomposition of the partonic structure to be made,
differential in exact parton kinematics. The hard-scattering has a
systematic perturbative expansion to all orders of perturbation
theory.  Instead of ordinary parton densities, the theorem uses parton
correlation functions.

Factorization also applies to the parton correlation functions, thus
enabling a systematic recursive decomposition of the fully exclusive
parton structure of generated events.  The formalism covers
arbitrarily nonleading logarithms with the use of an expansion of the
coefficient functions in powers of the coupling, without logarithms,
and with the use of suitable renormalization-group transformations.
It lends itself very naturally to a recursive implementation suitable
for a MCEG.  

The new factorization theorems that we have just summarized give a
decomposition of the final-state into its exclusive components.  We
also need separate factorization theorems that give the cross section
and the parton correlation function integrated over final states;
these enable the first pair of factorization theorems to provide
conditional probabilities to be used for the MCEG.  For the cross
section integrated over all final states, factorization is just
ordinary factorization, which expresses it in 
terms of ordinary parton densities.  We then showed that an almost
identical argument gives a very similar result for the parton
correlation function.

The formalism is completed with the aid of renormalization group
and DGLAP evolution, that allow perturbative calculations to be done
at the appropriate scales in each component of the different
factorizations.  The same ideas apply to final-state showering, but in
a simpler way \cite{phi3}.

Using the subtractive procedure proposed in \cite{phi3} to define
higher order corrections to the kernels, we constructed an algorithm
for a MCEG for DIS (including initial-state parton showering), in
which coefficient functions for both the hard scattering and the
showering can be obtained at arbitrarily non-leading order.  The
subtractions are applied point-by-point in momentum space, so that the
hard scattering coefficients and the evolution kernels are normal
functions, unlike the singular generalized functions that appear in
inclusive calculations from conventional factorization.

Because of the subtractions, the hard scattering coefficients are not
necessarily positive.  An EG with weights is in general to be expected
\cite{NLO.MC}.  An algorithm for weighted EG was proposed in
\cite{thesis}.

Although our derivations only apply to non-gauge theories, and there
is much work to be done to extend our work to full QCD, we believe our
work indicates how one should approach the full QCD problem.  One
considers the regions of momentum space that dominate but without
relying on any unitarity cancellation of a soft region.  This is like
the situation in factorization with transverse-momentum-dependent
(TMD) parton densities \cite{TMD,ji}, which work applies to inclusive
cross sections that are sensitive to parton transverse momentum.  In
QCD (and any gauge theory), two complications arise.  The first is
that there is an extra non-canceling soft factor to be treated.  The
second is that to be gauge-invariant, the parton correlation functions
must be defined with Wilson lines in the operators, and factorization
requires \cite{ji,CH1,CH2,pdf,metz} the Wilson lines to be in certain
directions.  Examination of the general techniques used in these works
suggests that they should continue to apply to the weighted cross
sections we use to understand the final state.  Naturally the
appropriate factorization must involve gauge-invariant unintegrated
parton correlation functions, so that parton momenta can be treated
exactly.  There are extra techniques already available for handling
the extra complications, for example the Collins-Soper evolution
equation for the direction dependence of the Wilson lines.  So many of
the elements needed to extend our results on MCEGs to full QCD are
available, and simply need to be adjusted and applied in the new
context.

Recently Collins and Metz \cite{metz} have performed a detailed
general analysis of the directions of the Wilson lines needed for
factorization; technically it depends on an understanding of the
allowed directions of certain contour deformations in virtual graphs
that are needed in the proof of factorization.  In the case of both
$e^+e^-$-annihilation and of DIS, their analysis of TMD factorization
does not have any requirement that the cross section be inclusive.
Since such cross sections involve a soft factor, no unitary
cancellations are needed to get factorization.  But one will probably
need multi-parton soft factors to accompany multi-parton production
processes, with corresponding new nonperturbative features.

It is the use of TMD factorization that saves the need for the unitary
cancellation of the soft factor that happens in non-TMD
factorization. 

But as Collins and Metz explain, TMD factorization in hadron-hadron
processes, like Drell-Yan, continues very critically to need a
unitarity cancellation.  Therefore, the limits to MCEGs are likely to
be more severe here.  This suggests that a full treatment will need a
discussion of spectator-spectator interactions that goes far beyond a
normal factorization framework.  Elements of this are undoubtedly
already present in some existing EGs like PYTHIA.

\acknowledgments One of us (X.~Zu) would like to thank Jeff Owens for
valuable discussions. This project is supported in part by the U.S. DOE.

%*************************************************************************

\end{document}